\documentclass[final]{ws-mpla}
\usepackage[super,compress]{cite}
\usepackage[breaklinks]{hyperref}  
\usepackage{graphicx}
\hypersetup{colorlinks,urlcolor=black,citecolor=black,linkcolor=black,filecolor=black}
\usepackage{xspace}

\newcommand{\mjj}{\ensuremath{m_\mathrm{jj}}\xspace}

\newcommand{\nsubj}{\ensuremath{\tau_{21}}\xspace}
\newcommand{\ddt}{\ensuremath{\tau_{21}^{DDT}}\xspace}

\newcommand{\VV}{\ensuremath{\PV\PV}\xspace}

\providecommand{\HERWIG} {{\textsc{herwig}}\xspace}

\providecommand{\MADGRAPH} {\textsc{MadGraph}\xspace}

\providecommand{\POWHEG} {{\textsc{powheg}}\xspace}

\providecommand{\PYTHIA} {{\textsc{pythia}}\xspace}

\newcommand{\TeV}{\ensuremath{\,\text{Te\hspace{-.08em}V}}\xspace}

\newcommand{\fbinv} {\mbox{\ensuremath{\,\text{fb}^\text{$-$1}}}\xspace}

\newcommand{\PT}{\ensuremath{p_{\mathrm{T}}}\xspace}
\newcommand{\pt}{\ensuremath{p_{\mathrm{T}}}\xspace}

\newcommand{\ptvecmiss}{\ensuremath{{\vec p}_{\mathrm{T}}^{\kern1pt\text{miss}}}\xspace}

\providecommand{\ttbar}{\ensuremath{\mathrm{t}\overline{\mathrm{t}}}\xspace}

\providecommand{\PV}{\ensuremath{\mathrm{V}}\xspace}

\providecommand{\PZpr}{\ensuremath{\mathrm{Z}^\prime}\xspace} 

\providecommand{\PWpr}{\ensuremath{\mathrm{W}^\prime}\xspace} 

\newcommand{\rpv}{\ensuremath{\rlap{\kern.2em/}R}\xspace}

\begin{document}

\thispagestyle{plain}

\def\bib{B\kern-.05em{I}\kern-.025em{B}\kern-.08em}
\def\btex{B\kern-.05em{I}\kern-.025em{B}\kern-.08em\TeX}
\newcommand{\sqrts}{\ensuremath{\sqrt{s}}}
\newcommand{\GEV}{\ensuremath{\rm{GeV}}}
\markboth{Thea Aarrestad}
{Searching for diboson resonances in the all-hadronic final state}

\catchline{}{}{}{}{}

\title{SEARCHING FOR DIBOSON RESONANCES\\
 IN THE BOOSTED ALL-HADRONIC FINAL STATE \\
 AT $\sqrt{\rm{s}}=13$~TeV WITH CMS
}

\author{\footnotesize THEA AARRESTAD}

\address{European Organization for Nuclear Research (CERN)\\
  CH-1211 Geneva 23, Switzerland\\
thea.aarrestad@cern.ch}

\maketitle


\begin{abstract}
This article summarizes three searches for diboson resonances in the all-hadronic final state using data collected at a center-of-mass energy of $\sqrt{\rm{s}}=13$~TeV with the CMS experiment at the CERN LHC. The boson decay products are contained in one large-radius jet, resulting in dijet final states which are resolved using jet substructure techniques. The analyses presented use 2.3 \fbinv, 35.9 \fbinv and 77.3 \fbinv of data collected between 2015 and 2017. These include the first search for diboson resonances in data collected at a 13 TeV collision energy, the introduction of a new algorithm to tag vector bosons in the context of analyzing the data collected in 2016, and the development of a novel multidimensional fit improving on the sensitivity of the previous search method with up to $30\%$. The results presented here are the most sensitive to date of all diboson resonance searches in the dijet final state. An emphasis on improvements in technique for vector boson tagging is made.

\end{abstract}


\section{Introduction}	
The Standard Model of particle physics (SM) is an extremely successful and predictive theory describing all the known elementary particles and how they interact through the electroweak and strong interaction. Despite its success, however, the SM cannot be the ultimate theory of particle physics due to its shortcomings in accommodating certain phenomena. One challenge is to comfortably fit the gravitational force into the SM framework at high energies (or, correspondingly, short distances). Gravity is beautifully described in General Relativity (GR) as a classical theory: a force caused by the curvature of space-time in the presence of matter and energy. The theory does not utilize quantum fields and energy is not quantized. The scales between the SM, a quantum field theory (QFT), and GR are completely different: space-time is curved on astronomical scales, where variations in space-time are essentially invisible. Hence, to the SM,
space-time is approximately flat and gravity does not exist. In order to have an
elegant unified theory of all the forces, attempts have been made to have a QFT
of the gravitational force by extending the SM particle family to incorporate a
particle to mediate the gravitational force called the graviton, a massless gauge
boson of spin-2. In the low-curvature and low-energy regime, the SM and GR are
fully compatible, and GR is an inevitable consequence of the quantum mechanics
of interacting gravitons. However, in the generic case, loop divergences that cannot
be reabsorbed through renormalization have hindered every effort of integrating
gravity in the SM in a renormalizable way. This has led to several proposals for
extending the SM in order to incorporate force mediating gravitons. One of these,
the Bulk Scenario of Warped Extra Dimensions~\cite{Randall_1999,PhysRevD.76.036006,Fitzpatrick_2007}, predicts TeV-scale gravitons, which preferably decay into vector bosons. In addition to the difficulties of incorporating gravity into a quantum field theory framework, problems occur at small
distances, of the order of the Planck scale, at which quantum gravitational effects
would become apparent. The Planck mass is $10^{16}$ times heavier than the W and Z
bosons, such that there is a hierarchy between the mass scales of gravity and the
electroweak forces. This is related to the value of the Higgs vacuum expectation
value (VEV), which gives the W and Z bosons their mass, and which predict much higher masses than those observed due to quantum loop corrections. Compositeness~\cite{Bellazzini:2014yua, Contino:2011np} solves the hierarchy problem by assuming that the SM breaks down at an
energy between the weak and Planck scales and that, around the TeV scale, the
Higgs boson no longer appears to be a single scalar particle but a composite state
of two fermions. This removes the hierarchy problem as there is no longer an ele-
mentary scalar in the SM and therefore no loop corrections that drive the mass to
the Planck mass. The compositeness theory is based on the breaking of two parallel
global symmetries such that three gauge bosons of the symmetry groups are predicted: $\mathrm{W}^{\prime\pm}$ and  $\mathrm{Z}^{\prime}$. These have masses of the order of the compositeness scale can have enhanced decays into bosons for certain choices of coupling parameters,
making searches for Beyond Standard Model (BSM) physics in diboson final states
well motivated. Several searches for resonant new particles decaying to WW, WZ or ZZ were performed using data collected with the CMS and ATLAS experiments at a center-of-mass energy of $\sqrt{\rm{s}}=8$~TeV, in Refs.~\protect\refcite{Khachatryan:2014hpa,Khachatryan:2014gha,Khachatryan:2014xja,Sirunyan:2016cao,Sirunyan:2018iff} by CMS and  Refs.~\protect\refcite{ATLASwprimeWZPAS,Aad:2014xka,Aad:2015ufa,Aad:2015owa,Aaboud:2017ahz,Aaboud:2017eta} by ATLAS. As SM vector bosons predominantly decay ($\sim68\%$ of the time)into quarks, searches in the all-hadronic final state become especially important when looking for resonances with low cross sections compared to SM backgrounds. How- ever, QCD multijet production is an overwhelming background and complicates any search in an all-jet final state. Further complicating these types of searches is that the diboson final states under consideration are challenging to resolve due to the bosons being highly energetic ('boosted'), resulting in the two quarks from the decay being collimated and merging into a single jet. This leads to a dijet fi- nal state topology where each jet has energy distributed bi-modally along the two quarks directions and is expected to have a mass around the W or Z boson mass. These qualities together, jet substructure and jet mass, can be used to discriminate quark- or gluon-initiated jets from W/Z jets, which is referred to as vector boson tagging (V-tagging). \vskip 1.em
\noindent In this article, the search for heavy resonances decaying to dibosons in the all-hadronic final state using data collected with the CMS experiment at $\sqrt{s}=13$~TeV will be discussed. This includes three different analyses performed between 2015 and 2019, with increasingly larger datasets and with a focus on novel V-tagging algorithms, which were developed and commissioned as part of the search effort. This article will present the first search for diboson resonances using data collected with a center-of-mass energy of 13 TeV, as published in Ref.~\protect\refcite{first13tev}. Then, the commissioning of a new V-tagging algorithm in the context of the analysis presented in Ref.~\protect\refcite{Sirunyan:2017acf} will be discussed and, finally, a novel framework for performing multidimensional searches, as introduced in Ref.~\protect\refcite{b2g18002}, will be summarized.

\section{The first search for diboson resonances at 13 TeV}
\label{sec:1}
In June 2015, as the CERN Large Hadron Collider began colliding protons again with an unprecedented collision energy of $\sqrt{s}=13$~TeV, the ATLAS Collaboration published an analysis of the full data set collected at $\sqrt{s}=8$~TeV in their search for heavy resonances decaying to dibosons. A significant deviation from the background expectations around 2 TeV was observed, where the largest deviation corresponded to a $3.4~\sigma$ local significance~\cite{Aad2015}.
The corresponding CMS analysis, documented a $1.3~\sigma$ excess at roughly the same resonance mass~\cite{Khachatryan:1700394}. The two measurements were found to be compatible, favoring a heavy resonance with a production cross section of around 5 fb and a mass between 1.9 and 2.0 TeV decaying to either WW, WZ or ZZ~\cite{Dias:2015mhm}. These measurements made it a priority to analyze the new 13 TeV data and look for a consistent signal. Due to the increase in center-of-mass energy from 8 to 13 TeV and a corresponding increase in partonic luminosity, one would expect the same number of signal events as were collected in the 20 $\rm{fb}^{-1}$ data set with a center-of-mass energy of 8 TeV, for a considerably smaller luminosity with a center-of-mass energy of 13 TeV, making it possible to confirm or exclude the excess within a few months of data taking. In the following, the first search for diboson resonances using data collected with $\sqrt{s}=13$~TeV in 2015 and first presented in Ref.~\protect\refcite{first13tev} will be summarized.

\subsection{General search strategy}
\begin{figure}[ht!]
\centerline{\includegraphics[width=0.79\textwidth]{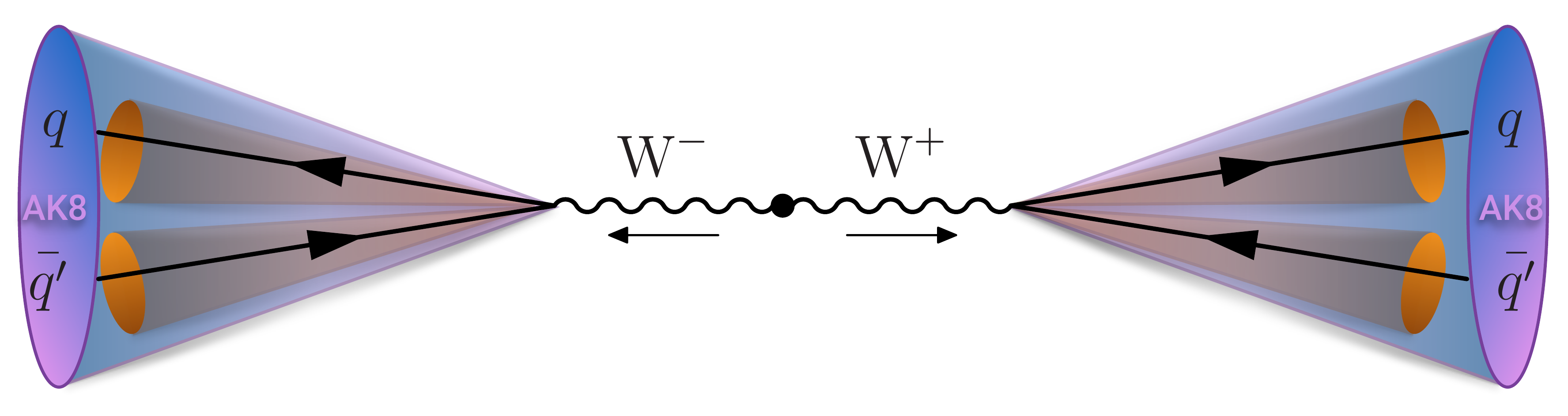}}
\vspace*{8pt}
\caption{If a heavy ($> 1$ TeV) resonance decays into vector bosons, the transverse momentum of each boson will be large and its decay products are merged into one single large cone AK8 jet.\protect\label{fig2}}
\end{figure}

The analysis strategy consists of selecting final states with two high-momentum, large-radius anti-$\rm{k}_{\rm{T}}$~\cite{Cacciari:2008gp} jets. As the distance between the two quarks $\rm{R_{qq}}\propto \rm{m_{V}/p_{T,V}}$, where $\rm{m_V}$ is the vector boson mass and $\rm{p_{T,V}}$ is its transverse momentum, the two quarks are expected to be merged into a single AK R=0.8 (AK8) jet above $\rm{p_{T,V}}>200~$GeV. This results in a dijet final state, as illustrated in Figure~\ref{fig2}.
These jets are expected to have a mass around the W or Z boson mass, and some intrinsic substructure stemming from their two-pronged decay. The invariant mass of the dijet system, $\rm{m}_{\rm{jj}}$, should be roughly equal to the resonance mass $M_X$, and is the parameter of interest. The main background for such an analysis consists of QCD multijet events since quark- and gluon-jets can obtain a high mass due to radiation from processes not coming from the primary hard scattering and, due to the large cross section of QCD processes, the number of QCD quark and gluon jets with a mass compatible with the W mass can be large. In order to discriminate between the two, three properties are considered. First, the {\it groomed mass} is used. Grooming was introduced as a tool to improve the mass resolution of large radius jets without significantly changing the background and signal event numbers. It consists of removing the softest parts of a jet in order to resolve its 'true' mass, by reclustering the jet and for every clustering step identify and remove soft particles consistent with radiation. Second, signal jets should appear two-prong like, as opposed to the single prong structure of quark and gluon jets. This is quantified through the N-subjettiness algorithm~\cite{Thaler:2010tr}, which attempts to count the number of hard sub-elements within a jet. Third, the dijet invariant mass for the signal process should peak around the resonance mass while the QCD spectrum is predicted to be smoothly falling. The strategy therefore, consists of performing a smoothness test on $\rm{m}_{\rm{jj}}$ of the observed data, a so-called “bump hunt”, by assuming that the signal will appear as a bump on top of a smooth distribution. The benefit of such a method is that a simulation of the background is not needed and so the procedure is robust. The disadvantage is that the analysis is intrinsically limited to regions where the dijet invariant mass spectrum is smooth, and hence regions with discontinuities due to trigger turn-ons or kinematic selections must be avoided.

\subsubsection{Jet substructure at trigger level}
Searches with boosted-jet final states in CMS has historically taken advantage of triggers based on $H_T$, which is the scalar sum of the reconstructed transverse jet energies. Due to an overwhelming QCD background in all-hadronic final states, the threshold for fully-hadronic triggers is very large in order to keep the trigger rate low (preferably around 10-30 Hz), resulting in $H_T$ thresholds between 0.7-1.1 TeV. This directly affects at which mass the search for resonances can begin since the background must be smoothly falling, requiring a trigger efficiency $>99\%$ in order to avoid turn-on effects, and, the full signal dijet invariant mass peak must be contained in order to get an accurate estimate of the signal shape. During the LHC shut down between \sqrts~8 and \sqrts~13 TeV data taking, new triggers that place requirements on the groomed mass of the jet were developed. Taking advantage of such 'boosted-jet' triggers in combination with $H_T$-based triggers, allows the dijet invariant mass threshold to be significantly lower, in this case by 75 GeV, allowing the search to start at dijet invariant masses of 1 TeV. These triggers were never before used in data, and this analysis was the first published result taking advantage of grooming at the trigger level in CMS.
\vskip 1.em
\noindent In addition to selecting the two highest-\PT jets in the event with $\PT>$200~GeV and requiring a dijet invariant mass of 1 TeV, a selection is made on the pseudo-rapidity between the two jets. The angular distribution between jets stemming from QCD multijet production, mainly t-channel scattering, is very different from the s-channel signal processes under study. The crossection for QCD t-channel processes as a function of the opening angle with respect to the beam axis $(\theta^*)$, exhibits a pole around $\cos{\theta^*}=1$, meaning QCD t-channel jets are mostly produced in the forward direction, with an opening angle with respect to the beam axis that is close to zero~\cite{HARRIS_2011}. The s-channel signal jets on the other hand, are more isotropic and hence concentrated in the barrel region. Jets are therefore required to have a separation of $|\Delta \eta|<1.3$.

\subsection{Vector boson tagging}
\begin{figure}[ht!]
\centerline{
\includegraphics[width=0.49\textwidth]{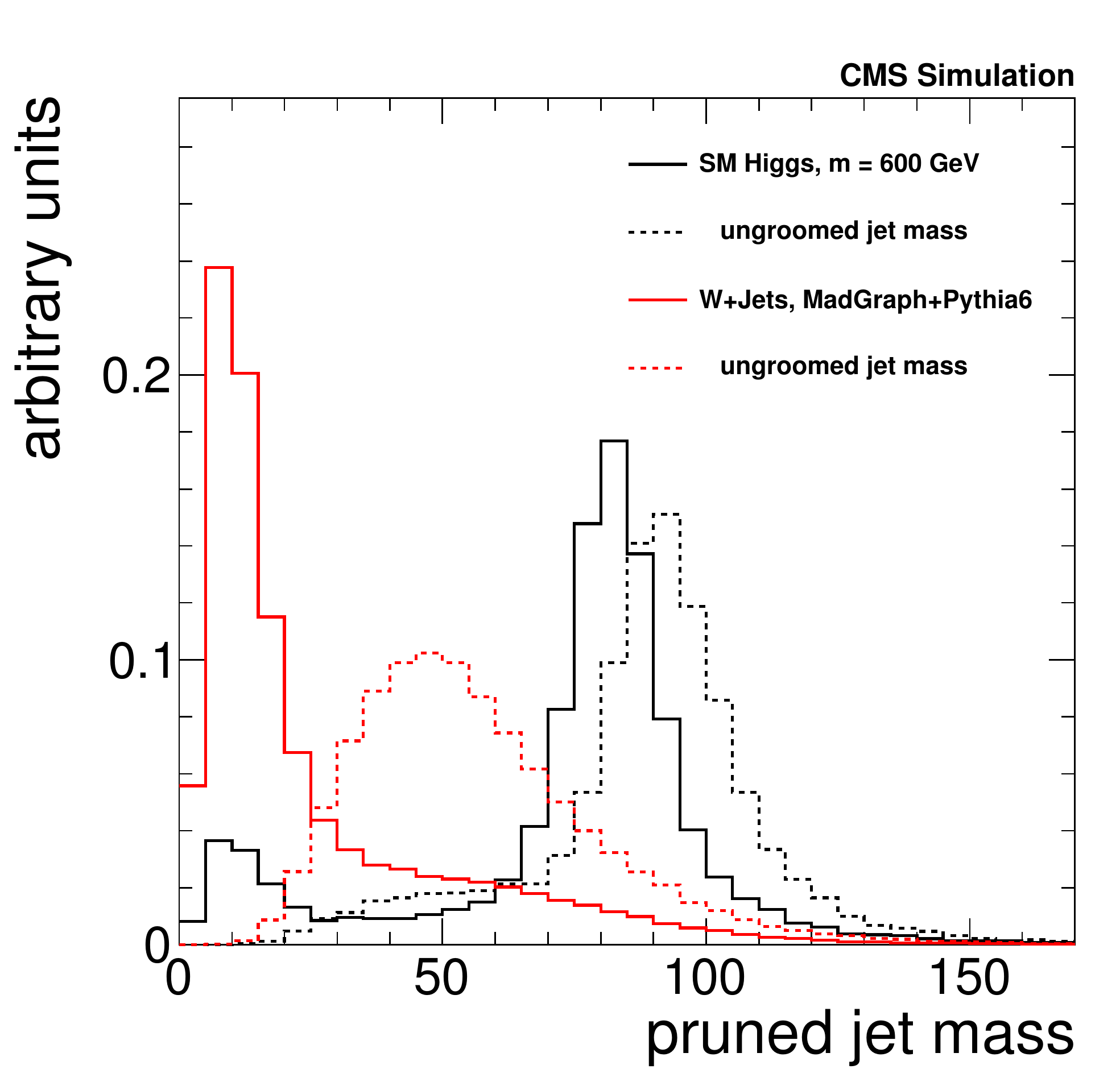}
\includegraphics[width=0.49\textwidth]{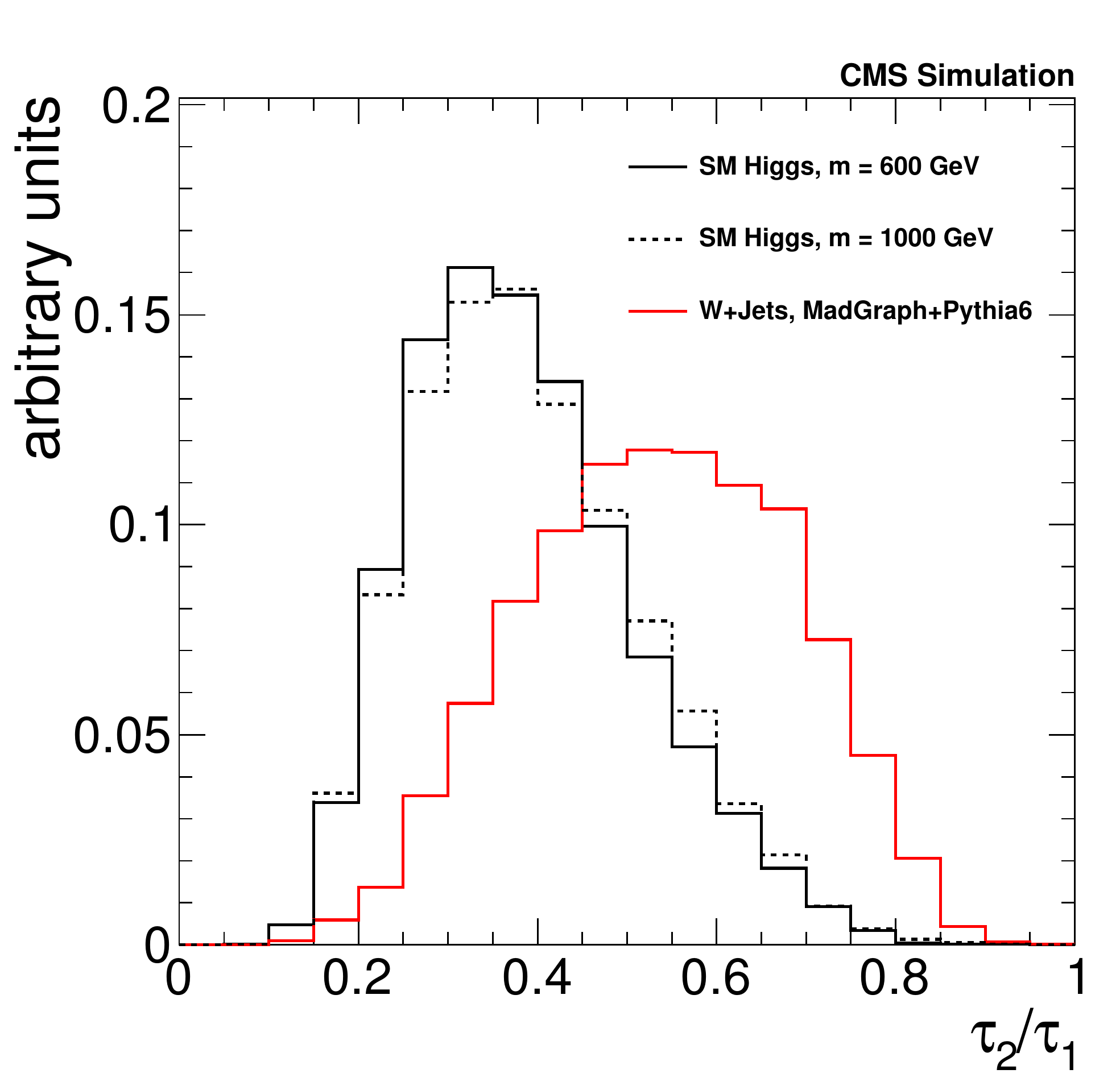}}
\vspace*{8pt}
\caption{Left: The jet mass distribution before (dotted line) and after (solid line) the pruning algorithm has been applied. Shown for merged W-boson jets coming from the decay of a heavy SM Higgs boson (black) and for quark- or gluon-jets in W+jets (red). Right: The N-subjettiness ratio \nsubj for merged W-boson jets coming from the decay of a heavy SM Higgs boson (black) and for quark- or gluon-jets in W+jets (red). Taken from Ref.~\protect\refcite{CMS-PAS-HIG-13-008}.}
\protect\label{fig3}
\end{figure}

As mentioned above, vector boson tagging consists of applying selections on the groomed mass and the N-subjettiness of the jet. The default grooming algorithm used by CMS for analyzing the data collected at 8 TeV, including for diboson resonance searches, was the “pruning” algorithm~\cite{PhysRevD.80.051501}. The pruning algorithm proceeds by re-clustering the jet with the Cambridge-Achen (C/A) algorithm~\cite{Dokshitzer:1997in}, requiring at each step involving constituents {\it a} and {\it b} that $\rm{min(p_{T,a},p_{T,b}) > z_{cut}p_{T,(a+b)}}$ and that the combination forms an angle smaller than $\rm{ D_{cut}}$ relative to the axis of recombination, where $\rm{ z_{cut}}$ and $\rm{ D_{cut}}$ are configurable parameters of the algorithm. If the criteria is not met, the softer of {\it a} and {\it b} is discarded. The reason for re-clustering using C/A, is that the clustering order is based solely on spatial separation so that the clustering history itself contains information about the presence of any geometrical substructure within a jet. Another grooming algorithm, the modified Mass Drop Tagger (mMDT)~\cite{Dasgupta:2013ihk}, generalized through the Soft Drop declustering method~\cite{Larkoski:2014wba}, was also under consideration but, due to it not yet having been studied in CMS data, was discarded in favor of the better understood pruning algorithm. The Soft Drop algorithm will be explored further in Section~\ref{sec:2}. Candidate vector boson jets are selected through a requirement on the pruned AK8 jet mass to be between 65 and 105 GeV, compatible with the W/Z mass, but avoiding overlaps with the Higgs boson mass region as well as avoiding the lower mass region dominated by quark- and gluon-jets from QCD multijet production. To further discriminate boson jets from quark or gluon jets, a selection on the jet N-subjettiness is applied. The N-subjettiness variable, $\tau_N$, yields a probability of the jet originating from $N$ prongs. Jets with $\tau_N=0$ have most of their constituents aligned along the subjet axes. However, if $\tau_N>>0$, a large fraction of the energy is radiated away from the subjet directions and the jet is more likely to have more than $N$ subjets. In CMS, and as recommended by the authors in Ref.~\protect\refcite{Thaler:2010tr}, the ratio $\tau_2/\tau_1$ (referred to as \nsubj) is used in order to discriminate W/Z jets from QCD jets. The reason for this is that, while signal jets are expected to have a large $\tau_1$, quark- or gluon-jets can similarly have large $\tau_1$ due to diffuse radiation present. However, QCD jets with a large $\tau_1$ tend to have an equally large $\tau_2$ whereas signal jets do not, hence the ratio of the two provides greater separation power. In CMS, the N-subjettiness algorithm is by default applied to ungroomed jets and a selection of $\nsubj<0.45$ is made. Such a selection on pruned jet mass and \nsubj together is referred to as V-tagging.  The vector boson signal efficiency when applying only the pruned jet mass selection is around $80\%$ with a quark/gluon-jet mistagging rate of $\sim15\%$. After additionally applying a \nsubj selection, the signal efficiency drops to around $55\%$ and the mistagging rate to $\sim 2\%$. 
The pruned jet mass (left) and \nsubj (right) distributions in Monte Carlo (MC) simulation are shown in Figure~\ref{fig3}. The signal jets are in this case fully merged W-jets stemming from the decay of a heavy Higgs (black), while the background jets are quark- or gluon-jets from a W+jet background sample (red). The jet mass is shown before (dotted lines) and after (solid lines) pruning is applied.
Differences in the spectra of V-tagging variables between data and simulation can cause differences in V-tagging efficiency, which are crucial to correct for in order to have an accurate estimate of the signal efficiency. This is done using a \ttbar-enriched region with a high content of boosted W-jets and will be explained in more detail in Section~\ref{sec:2}.

\subsection{Enhancing sensitivity with event categorization}
Events are classified into six categories according to their purity and mass. First, two selections on \nsubj are made: a {\it high purity (HP)} selection requiring $\nsubj<0.45$ and a {\it low purity (LP)} selection, requiring $0.45<\nsubj<0.75$. Then two purity categories are defined: The HPHP category, which has the highest signal significance, requires both jets to pass the HP selection and the HPLP category, where one jet is required to pass the HP selection and one is required to pass the LP selection. While the first category is most sensitive where the background is high, the latter category becomes important at very high resonance masses, or correspondingly high dijet invariant mass, where the background contribution is small and maximizing signal efficiency is therefore more important. Including this category improves the analysis above dijet invariant masses of around 3 TeV. Further, the pruned jet mass ($\rm{m_P}$) window is split into a W and a Z boson mass window where the W boson window is defined as $65~\GEV<m_P<85~\GEV$ and the Z boson window as  $85~\GEV<m_P<105~\GEV$. The benefit of this is that event counting in these windows can be used to determine whether a potential signal observation is due to the decay of a heavy resonance into WW, WZ or ZZ. For instance, the yield in the WZ category will be higher for a resonance decaying to WZ than one decaying to WW/ZZ. 
In addition, the combination of two mass categories leads to slightly better ($10\%$) or similar sensitivity as when using one large mass window. In total, this results in six different analysis categories, all of which are combined during statistical analysis.

\subsection{Background modeling}
The QCD multijets background is assumed to be described by a smooth, monotonically decreasing function, which is parametrizable. This is similar to what is done in previous CMS analyses looking for bumps in the dijet invariant mass spectrum~\cite{Chatrchyan:2012ypy,CMS-PAS-EXO-12-059}. The search is then performed by fitting the sum of the functions for background and signal to the dijet invariant mass spectrum in data. Neither data control regions nor simulated samples are used directly by this method. The background functions are leveled exponentials of the following form:
\begin{equation}
\label{eq:dijet2}
\frac{dN}{d\mjj}= \frac{ P_0(1-\mjj/\sqrt{s})^{P_1} } { (\mjj/\sqrt{s})^{P_2+ P_3 \ln(\mjj/\sqrt{s}) + P_4 \ln^2(\mjj/\sqrt{s})+ P_5\rm{...}} }\:\:,
\end{equation}
where $m$ is the dijet invariant mass, $\sqrt{s}$ is the center-of-mass energy, $P_0$ is a normalization parameter for the probability density function, and $P_{N,N>0}$ are shape parameters. The number of needed shape parameters depends on the category (more statistics usually require a higher number of parameters in order to describe the observed spectrum). Functions of this form have been utilized in dijet searches at hadron colliders for decades and work well due to two main reasons: The numerator is representative of the behavior of parton distribution functions and the first term in the denominator represents the lowest order behavior of matrix elements (mass to a power). All parameters are left floating in the simultaneous signal- and background-fit, effectively finding the lowest background shape possible that can accommodate the highest possible signal (at a given confidence level). The number of fit parameters is decided through a Fisher's F-test~\cite{RePEc:bla:istatr:v:80:y:2012:i:3:p:491-491}. In this test, starting from a 2-parameter function the goodness of fit (quantified through the sum of residuals) is compared when fitting the data signal region with a 3, 4 and 5 parameter function. The model with more parameters will always fit the data at least as well as the model with less parameters, but the F statistic defines whether the higher order model yields a significantly better fit to the data, quantified through a confidence level of the simpler function being correct (here CL = 10\%). 
The fit range is then chosen such that it starts where the trigger efficiency has reached its plateau to avoid bias from trigger inefficiency, and extends to the bin after the highest dijet invariant mass point in each category. The binning chosen for the fit follows the detector resolution as in Refs.~\protect\refcite{RePEc:bla:istatr:v:80:y:2012:i:3:p:491-491}, and is referred to as the 'dijet binning'. The uncertainty on the background parametrization is statistical only and is taken as the covariance matrix of the dijet fit function. In addition, different background parameterizations are studied and found to be within the fit uncertainty of the nominal fit.

\subsection{Signal modeling}
\begin{figure}[ht]
\centerline{\includegraphics[width=0.49\textwidth]{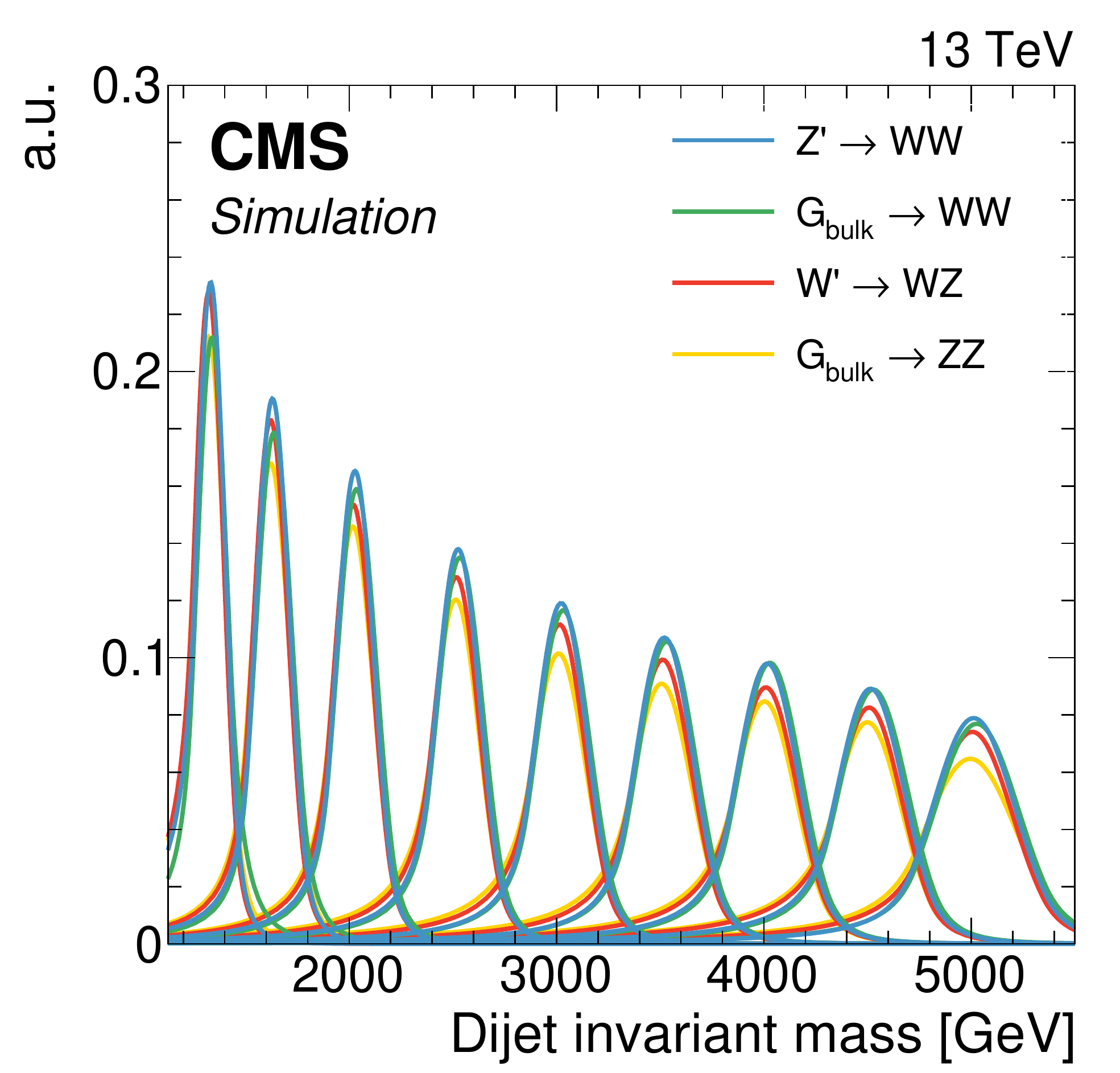}}
\vspace*{8pt}
\caption{Dijet invariant mass distribution for different mass and signal hypotheses. Taken from Ref.~\protect\refcite{b2g18002}.\protect\label{fig4} }
\end{figure}

The signal shape is extracted from signal MC with resonance masses in the range from 1 to 4 TeV. A linear interpolation provides shapes for the mass points in between in steps of 100 GeV. From these shapes, signal-shape models are constructed as composite models with a Gaussian core, due to detector resolution effects, and an exponential tail, due to radiation and the effects of parton distribution functions. The latter is due to the parton luminosities being bigger at low mass than at the resonance mass, especially at higher dijet mass where the parton-distribution functions rapidly decrease~\cite{HARRIS_2011}. Uncertainties in parameter shape due to uncertainties in jet energy scale and resolution are inserted by variations of the Gaussian peak position and width. The dijet invariant mass shape for different benchmark model signals is shown in Figure~\ref{fig4}. The signal and background components are then simultaneously fitted to the data points.
\subsection{Results}
The main systematic uncertainties in these types of searches is related to the signal modeling. Dominant of these is the uncertainty on the vector boson tagging efficiency scalefactor, which includes the statistical uncertainty, uncertainties due to simulation of the \ttbar topology used to extract the scale factor, and an extrapolation uncertainty of the scale factor to high \PT. Systematic uncertainties due to uncertainties in jet mass scale and resolution are inserted by scaling the jet mass up and down within uncertainties obtained in a \ttbar control region (described in Section~\ref{sec:wtagsf}). Uncertainties related to jet mass and \nsubj are correlated among analysis categories (e.g events migrating out of the HP category, move into the LP category). 

\begin{figure}[ht!]
\centering
\centerline{\includegraphics[width=0.49\textwidth]{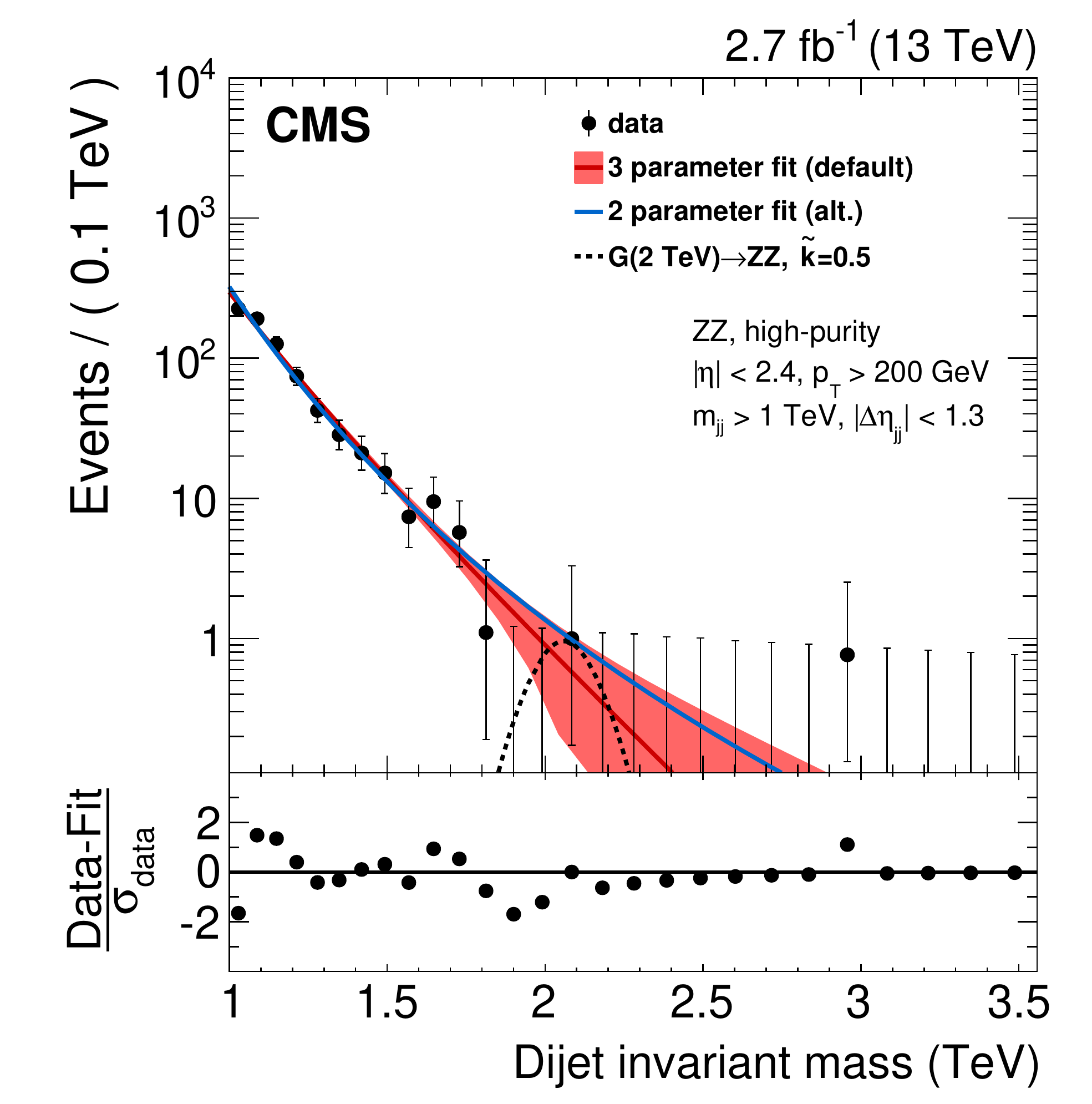}
\includegraphics[width=0.49\textwidth]{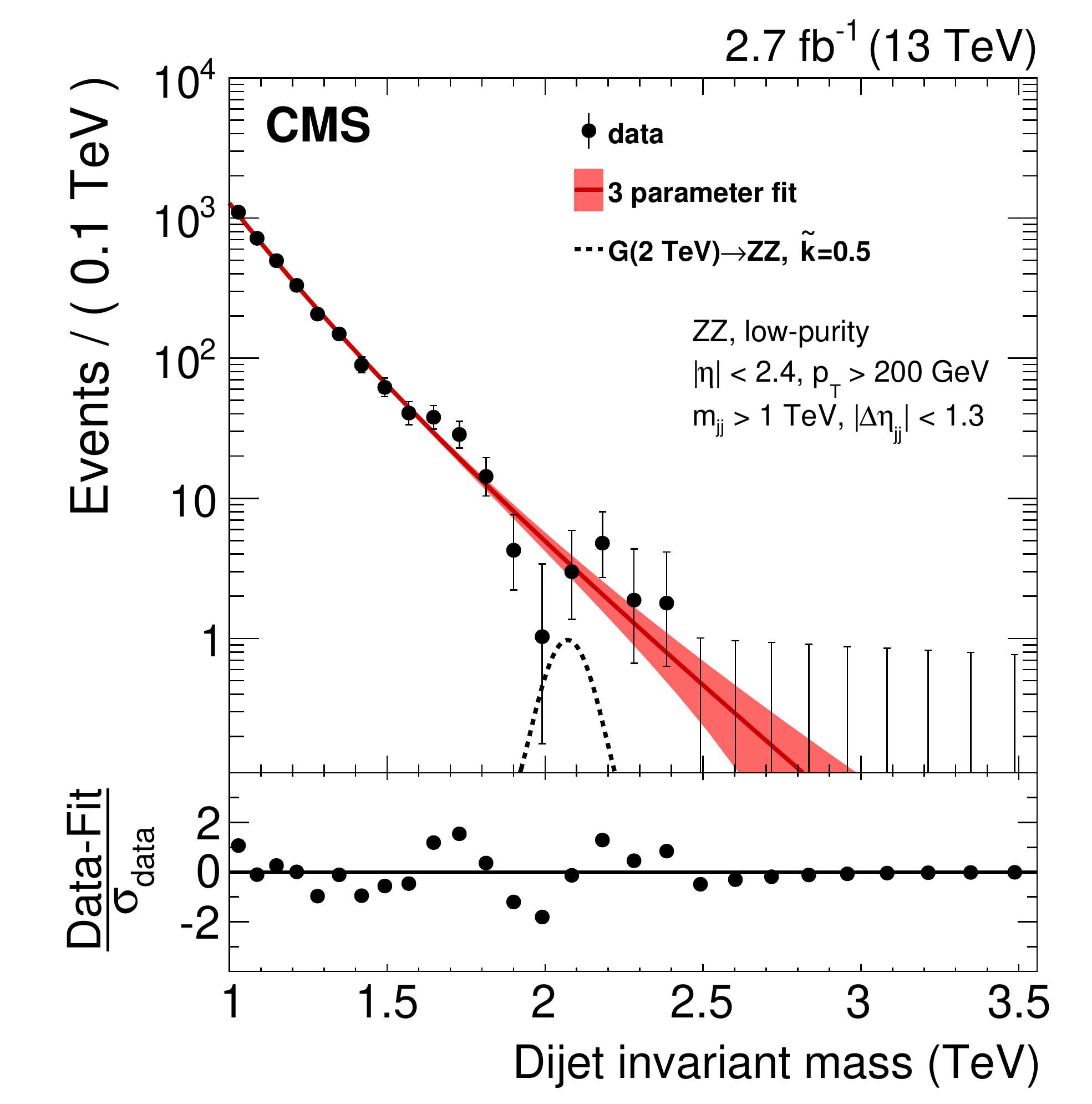}}
\vspace*{8pt}
\caption{Fit to data in the signal region using a background only fit for the ZZHP (left) and ZZLP (right) categories. The filled red area correspond to the $1 \sigma$  statistical error of the fit. In the fit to data in the ZZHP category (left) both the nominal (red) and an alternative (blue) fit is shown. Taken from Ref.~\protect\refcite{first13tev}.}
\protect\label{fig:search1:bkgfitMassCat}
\end{figure}

The background fits for two analysis categories in the data signal region for 2.7 \fbinv of data collected in 2015 are shown in Figure~ \ref{fig:search1:bkgfitMassCat}. Here, a background-only fit is performed while, as described above, a simultaneous fit is used for the limit-setting procedure. The filled area corresponds to the $1 \sigma$ error band of the background fit, obtained using linear error propagation. For the ZZHP category (left), both the nominal (red) and an alternative (blue) fit is shown, illustrating that the alternate fit is covered by the fit error band. This illustrates how a lack of constraint on the fit in the dijet invariant mass tail when statistics are very low can be a drawback of a method relying fully on a parametric fit and reduces the analysis sensitivity in the high dijet invariant mass region. This improves with integrated luminosity, resulting in more data points in the high mass tail which further constrain the fit. 
No excess is observed and upper limits on the cross section times branching ratio of the process $\text{X} \to \VV$ are set, using the asymptotic $\textrm{CL}_\textrm{S}$ method~\cite{CLs}. The binned likelihood is defined as
\begin{equation}
L = \prod_i\frac{\mu^{n_i}_ie^{-\mu_i}}{n_i!},
\end{equation}
with
\begin{equation}
\mu_i=\sigma \cdot N_i(S)+N_i(B).
\end{equation}
Here $\sigma$ is the signal strength scaling the expected number of signal events in the $i$-th dijet invariant-mass bin $N_i(S)$, $N_i(B)$ is the expected number of background events in dijet invariant-mass bin $i$ and $n_i$ is the observed number of events in the $ith$ dijet invariant-mass bin. The background per bin $N_i(B)$ is estimated from the background component of the best signal and background fit to the data points, with the signal cross section set to zero. The number of signal events in the $i$-th dijet invariant mass bin, $N_i(S)$, is estimated from the signal templates, considering only dijet invariant masses in a 20\% window around the resonance mass, containing most of the signal contribution while making sure to keep a good description of the core. To maximize the search sensitivity, the results are combined with those of the corresponding semi-leptonic analysis, as presented in Ref.~\protect\refcite{Sirunyan:2017acf}.
The all-hadronic analysis sets stronger upper limits than the semi-leptonic analysis above 1.7\TeV for a \PZpr and above 1.3\TeV for a \PWpr due to the higher hadronic branching fraction of W and Z bosons.
The combined results just excluded a $\PWpr$ with a mass around 2 \TeV, the favored candidate to explain the diboson excess observed in 8 TeV data. However, Bulk Graviton signals are still far from excluded and, with the expected increase in luminosity of a factor of ten in 2016, a follow-up analysis with improved taggers would be necessary to resolve the matter.

\section{Towards robust boosted jet tagging}
\label{sec:2}
\begin{figure}[ht!]
\centerline{\includegraphics[width=0.69\textwidth]{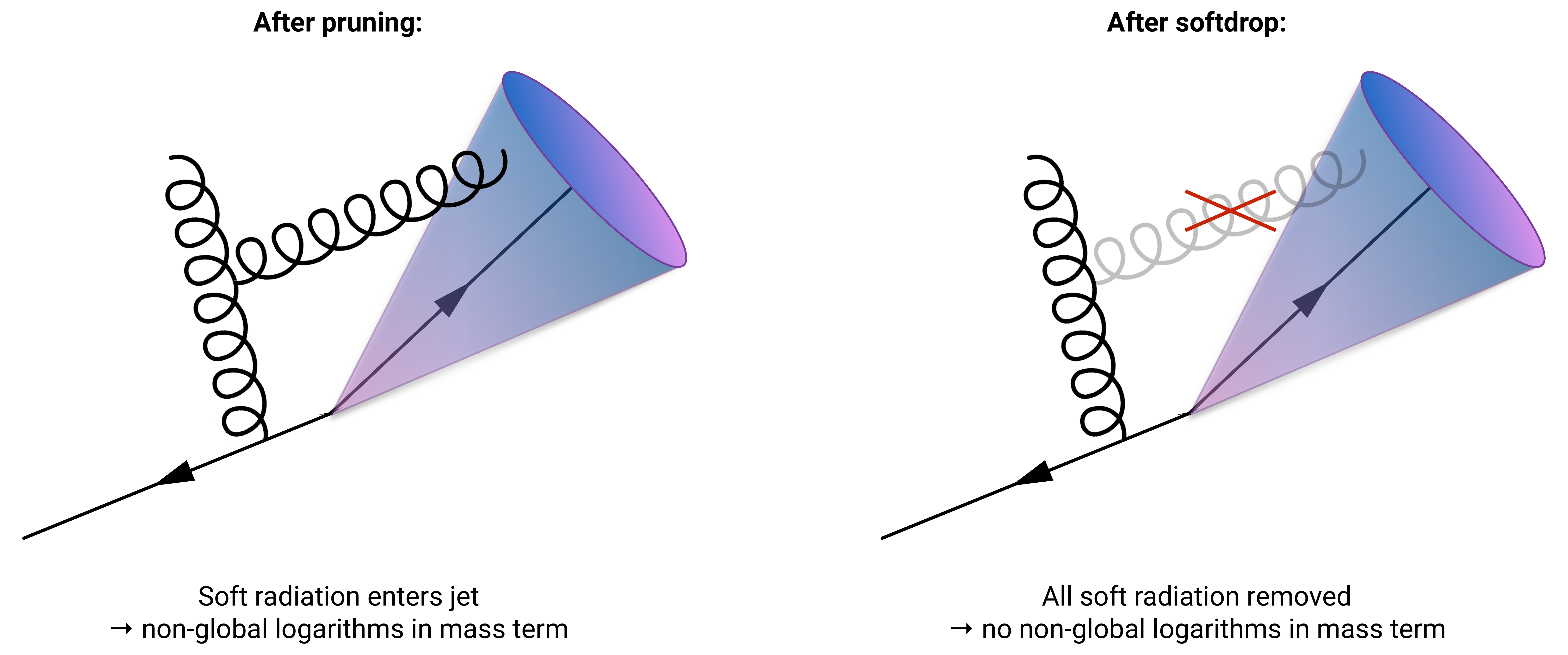}}
\vspace*{8pt}
\caption{The pruning algorithm does not remove all soft emission and therefore has non-global logarithmic terms in the jet mass. Softdrop with $\beta = 0$ removes all soft emissions wider than the dominant two-prong substructure and is therefore free of non-global logarithms.\protect\label{fig:ngls}}
\end{figure}

For proton-proton collisions starting in 2016, the LHC reduced the beta function at the collision point controlling how narrow, or “squeezed”, the beam is ($\beta^*$) from 80 to 40 cm, resulting in an instantaneous luminosity three times that of the peak luminosity in 2015. For CMS, this would lead to a large increase in collected data (eventually a factor of ten), but at the cost of a higher mean number of interactions per proton bunch crossing, causing additional interaction vertices per event (pileup) whose effect would need to be mitigated in order to maintain analysis sensitivity. A novel pileup-subtraction algorithm was proposed to deal with increasing pileup called Pileup per particle identification (PUPPI)~\cite{Bertolini2014}. PUPPI considers not only charged pileup, as is usually done in CMS through the {\it charged hadron subtraction (CHS)} algorithm, but also reweights neutral particles in the jet with their probability of arising from pileup. PUPPI has proven itself far superior to the existing pileup-removal algorithm in terms of jet observables, like the jet mass, for large radius jets.  As substructure variables are extremely sensitive to additional particles present in the event, having a V-tagging algorithm based on PUPPI jets is therefore desirable. In addition, the mMDT, or Soft Drop algorithm with $\beta = 0$, hereafter just referred to as "softdrop", was studied as an alternative to the pruning algorithm for its improved theoretical qualities: In addition to removing sensitivity to the soft divergences of QCD as all grooming algorithms do, namely they are {\it infrared and collinear (IRC) safe}, the softdrop algorithm eliminates all correlated soft emissions in the jet which are wider than the dominant two-prong substructure, emissions which lead to non-global logarithmic terms (NGLs) in the jet mass~\cite{Dasgupta:2013ihk}. NGLs arise from configurations where, for instance, a soft gluon is radiated into the jet cone, as illustrated in Figure~\ref{fig:ngls}. 
Being free of NGLs allows one to calculate the softdrop jet mass to a significantly higher precision than what is possible for other grooming algorithms or for the plain jet mass (NGLs are the main reason a full re-summation of the plain jet mass beyond next-to-leading-log accuracy does not exist). The focus of this search would therefore be on the commissioning of a novel vector boson tagging algorithm based on PUPPI and softdrop.

\subsection{A new pileup resistant and perturbative safe tagger}
 When studying the softdrop jet-mass of W-jets in 13 TeV simulation~\cite{CMS-PAS-JME-16-003}, a significant dependence on the jet transverse momentum and pseudo-rapidity was observed, both for reconstructed and particle-level jets. This shift was enhanced when applying standard CMS jet energy corrections.
 \begin{figure}[ht!]
\centerline{\includegraphics[width=0.99\textwidth]{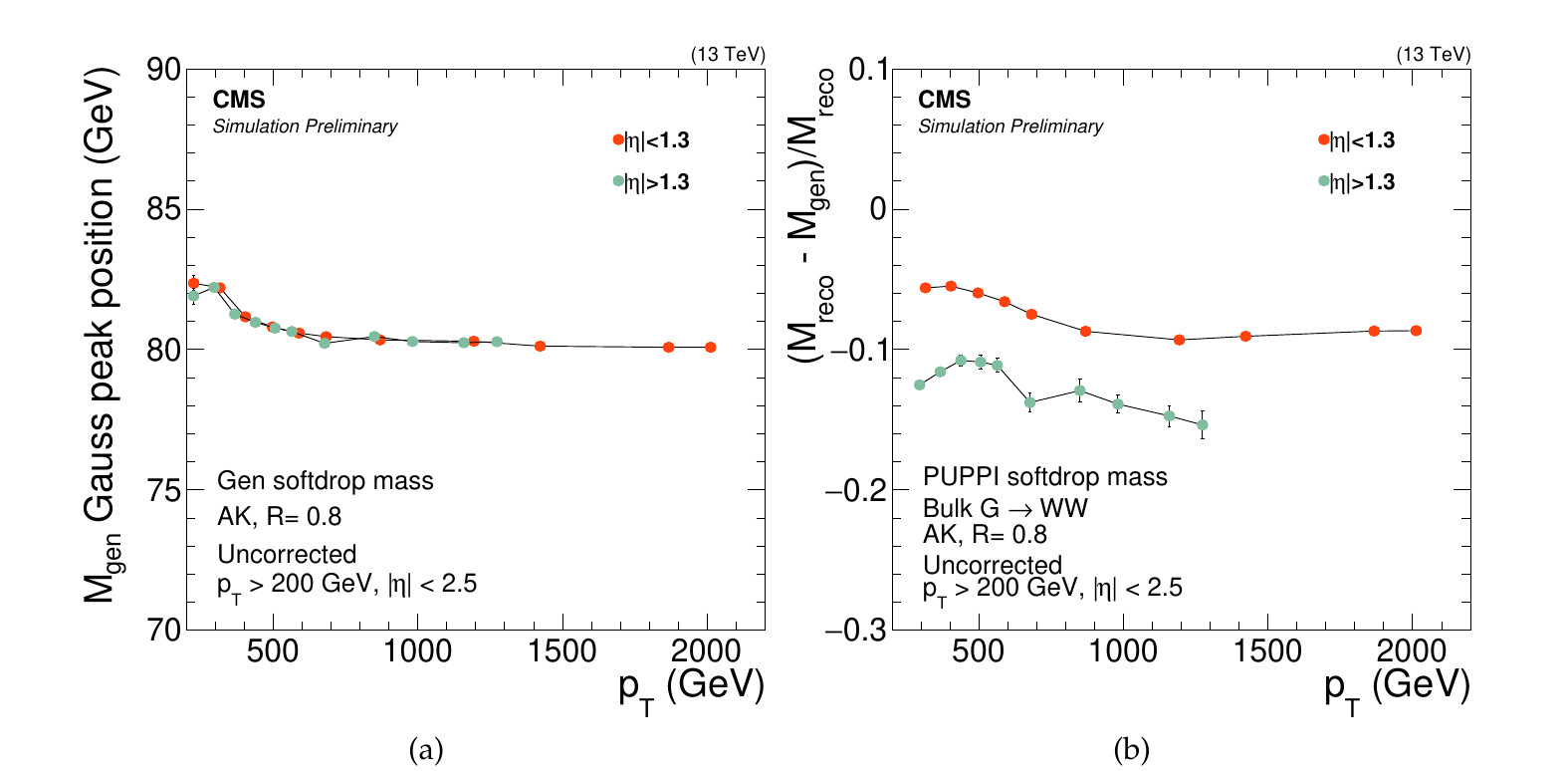}}
\vspace*{8pt}
\caption{(a) The Gaussian peak position of the fitted generator level W-jet softdrop mass distribution as a function of jet \PT and (b) the difference in reconstructed PUPPI softdrop jet mass and generated softdrop jet mass, as a function of jet \PT and in two bins of pseudo-rapidity. Taken from Ref.~\protect\refcite{CMS-PAS-JME-16-003}.\protect\label{fig:sdcorr}}
\end{figure}
Figure~\ref{fig:sdcorr} shows the mean of a Gaussian fit to the uncorrected PUPPI softdrop mass as a function of jet $\pt$ in two different $\eta$ bins (smaller or greater than $|\eta|=1.3$) for W-jets coming from a Bulk Graviton signal sample. A mass shift both as a function of $\eta$ and \PT is observed and the reconstructed W boson mass is significantly lower than the generated mass of 80 GeV.

\begin{figure}[ht!]
\centerline{
\includegraphics[width=0.59\textwidth]{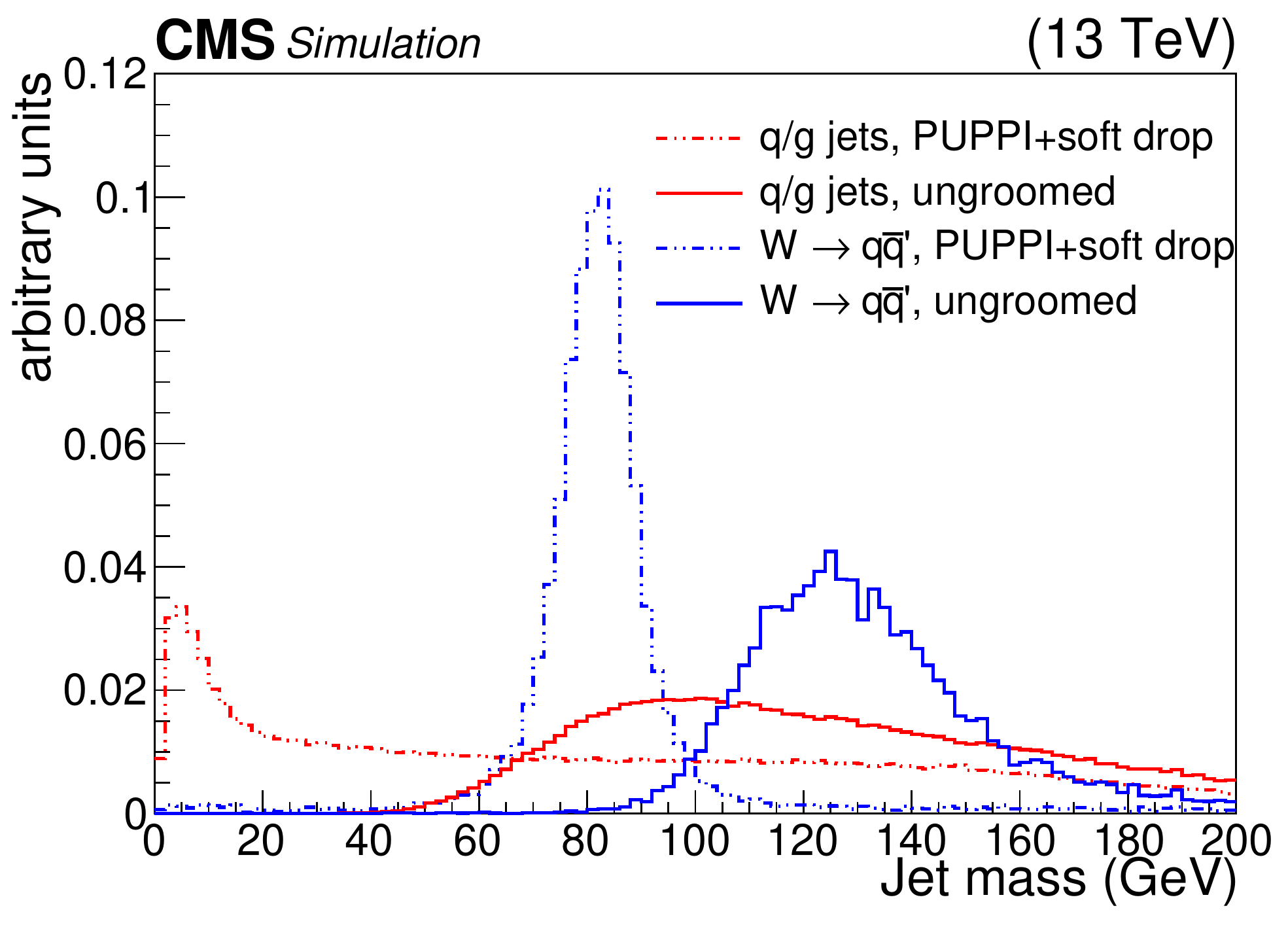}}
\vspace*{8pt}
\caption{The jet mass spectra for quark/gluon-jets (red) and W-jets (blue) before (solid line) and after (dotted line) softdrop has been applied. Taken from Ref.~\protect\refcite{Sirunyan:2017acf}.\protect\label{fig:PSD}}
\end{figure}

\begin{figure}[ht!]
\centerline{
\includegraphics[width=0.49\textwidth]{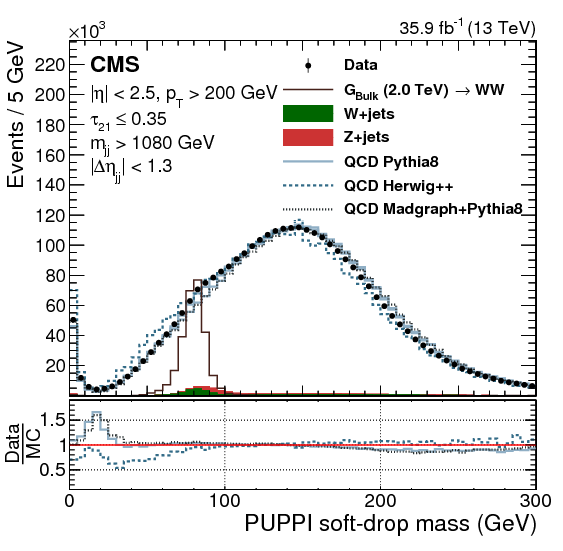}
\includegraphics[width=0.49\textwidth]{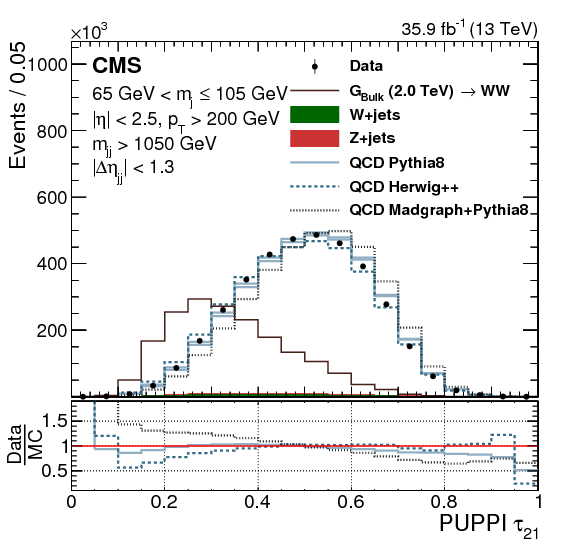}
}
\vspace*{8pt}
\caption{The PUPPI softdrop jet mass after a selection on PUPPI \nsubj (left) and the PUPPI \nsubj distribution after a selection on PUPPI softdrop mass. Taken from Ref.~\protect\refcite{Sirunyan:2017acf}.\protect\label{fig:puppisd}}
\end{figure}

\begin{figure}[ht!]
\centerline{
\includegraphics[width=0.49\textwidth]{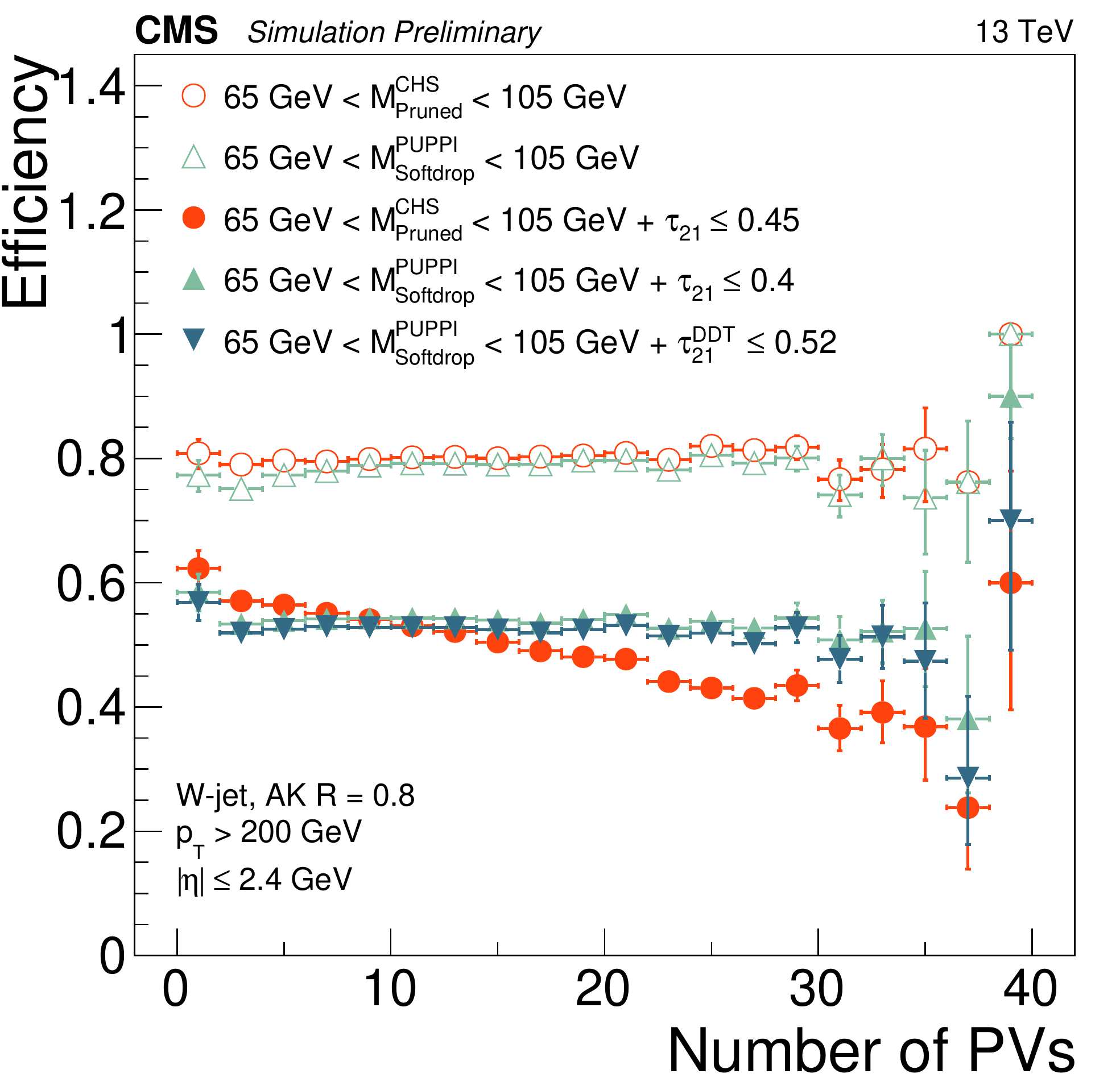}
\includegraphics[width=0.49\textwidth]{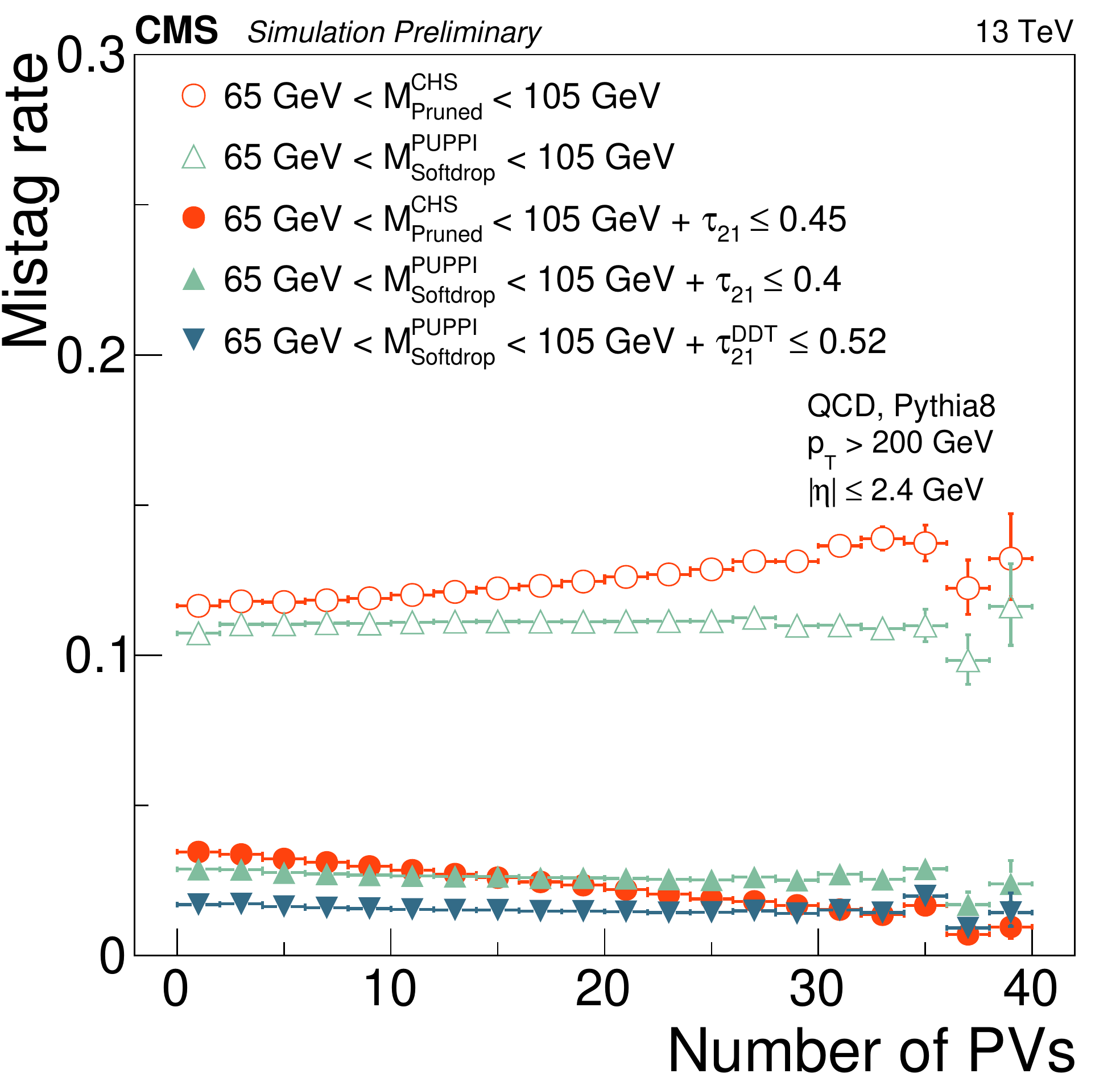}
}
\vspace*{8pt}
\caption{W-jet efficiency (left) and quark/gluon-jet mistagging rate (right) as a function of number of primary vertices in event.}
\protect\label{fig:effs}
\end{figure}

This is problematic as it results in a signal efficiency that varies with energy. For the pruned jet mass, such a dependence was not observed. The standard energy corrections are developed for AK R = 0.8 jets and are not guaranteed to be appropriate for softdropped or pruned jets. Therefore, dedicated softdrop jet mass corrections to account for an observed shift in jet mass scale as a function of jet\PT and $\eta$ were developed, starting from the observed spectrum in Fig.~\ref{fig:sdcorr}. These corrections are applied to all jets, both in data and simulation, and act as a correction of the jet mass scale. The jet mass spectrum before (solid line) and after (dotted line) PUPPI softdrop with softdrop corrections is applied, is shown in Fig.~\ref{fig:PSD} for quark/gluon jets (red) and W jets (blue). The corrected W-jet softdrop mass is now peaking at 80 GeV, in contrast to what was observed for the uncorrected softdrop mass in Fig.~\ref{fig:sdcorr}, and the quark/gluon jet softdrop mass is concentrated around zero at the true q/g mass. 
The PUPPI softdrop jet mass, after a selection of $\nsubj < 0.35$, and PUPPI \nsubj in 2016 data (black markers) and in simulation for three different MC generators (blue lines) is shown in Fig.~\ref{fig:puppisd}.
A strong dependence of substructure variables on shower generator is observed and the signal efficiency is therefore measured in a region better described by simulation. This is done in a dedicated measurement, explained in Sec.~\ref{sec:wtagsf}. In this measurement, any residual differences between data and MC with respect to the jet mass scale and resolution are also accounted for. The QCD jet mass spectrum is severely sculpted after a selection on \nsubj, and is now peaking around the signal mass pole rather than remaining smoothly falling as in Fig.~\ref{fig:PSD}. The jet mass spectrum is not fitted in this analysis, and therefore mass sculpting effects are not problematic. In Section~\ref{sec:4} this will become important, however, and methods to avoid mass sculpting will be discussed. The W-jet tagging efficiency (left) and quark or gluon jet mistagging rate (right) as a function of the number of primary vertices in the event (NPV) is shown in Figure~\ref{fig:effs}. When applying a mass selection only (hollow markers), the signal efficiency of a CHS pruned jet mass selection stays constant as a function of pileup, however, the mistagging rate increases by up to 30\% at an average pileup of 50 interaction vertices per event. For a selection on PUPPI softdrop jet mass, both efficiency and mistag rate stays constant. The signal efficiency for a mass cut only, is around 80\%. When additionally applying a \nsubj selection, the dependence on pileup is significant for a V-tagging algorithm combining CHS \nsubj with a CHS pruned jet mass selection. This underlines the strong dependence of substructure variables on pileup. For a selection on PUPPI \nsubj and PUPPI softdrop jet mass, the efficiency and mistagging rate is constant up to a pileup of 50.
The additional curves on the plot (\ddt) is for a mass-decorrelated version of \nsubj, which will be described further in Section~\ref{sec:3}. Using optimized selections on each tagger, a signal efficiency of $\sim55\%$ for a $\sim 2\%$ mistagging rate is observed for both taggers, assuming an average pileup of $\sim 30$ interaction vertices per event, as was the expected pileup in 2016. Detailed studies can be found in Ref.~\protect\refcite{CMS-PAS-JME-16-003}. The resilience of the PUPPI algorithm to pileup and the theoretical robustness of the softdrop algorithm, motivated a switch to a PUPPI softdrop based V-tagging algorithm, and this became the default V-tagging algorithm in CMS in 2016 (together with the dedicated softdrop jet mass corrections mentioned above).

\subsection{Correcting V-tagging efficiency}
\label{sec:wtagsf}
\begin{figure}[ht]
\centerline{\includegraphics[width=0.49\textwidth]{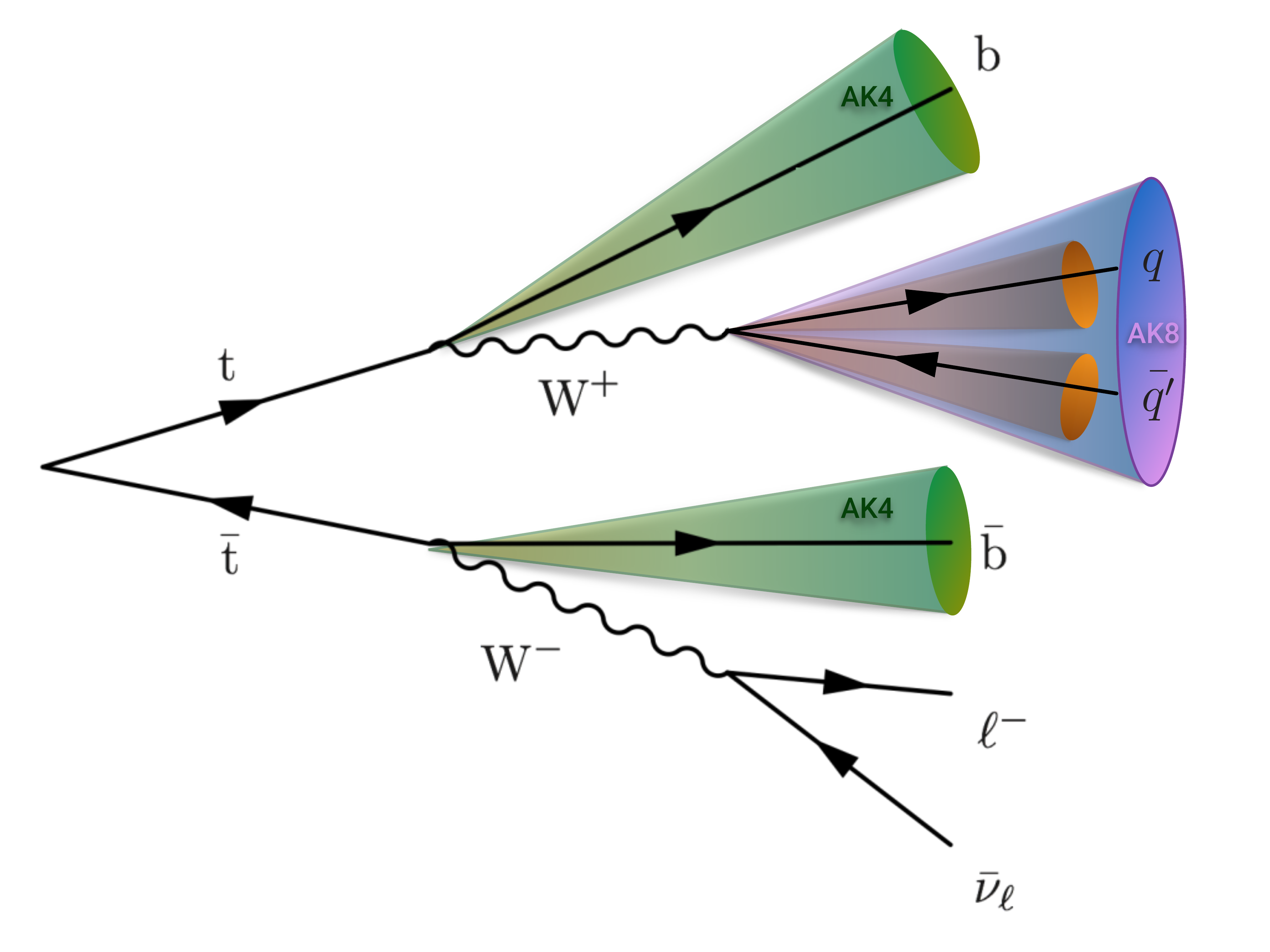}}
\vspace*{8pt}
\caption{A \ttbar-region enriched in merged W-jets is used to measure V-tagging efficiency, jet mass scale and resolution.\protect\label{fig:sltt}}
\end{figure}

\begin{figure}[ht!]
\centerline{\includegraphics[width=0.49\textwidth]{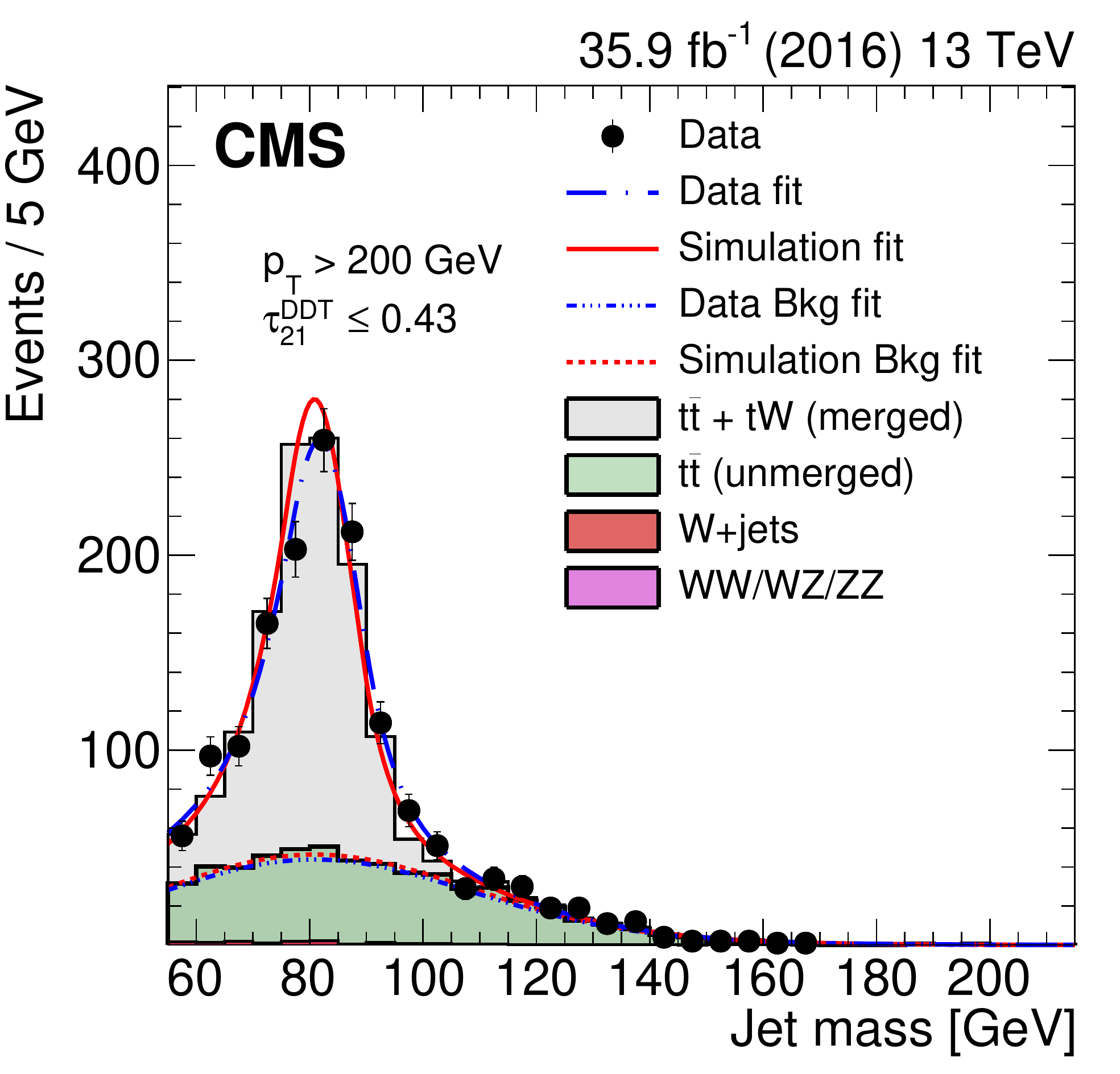}
\includegraphics[width=0.52\textwidth]{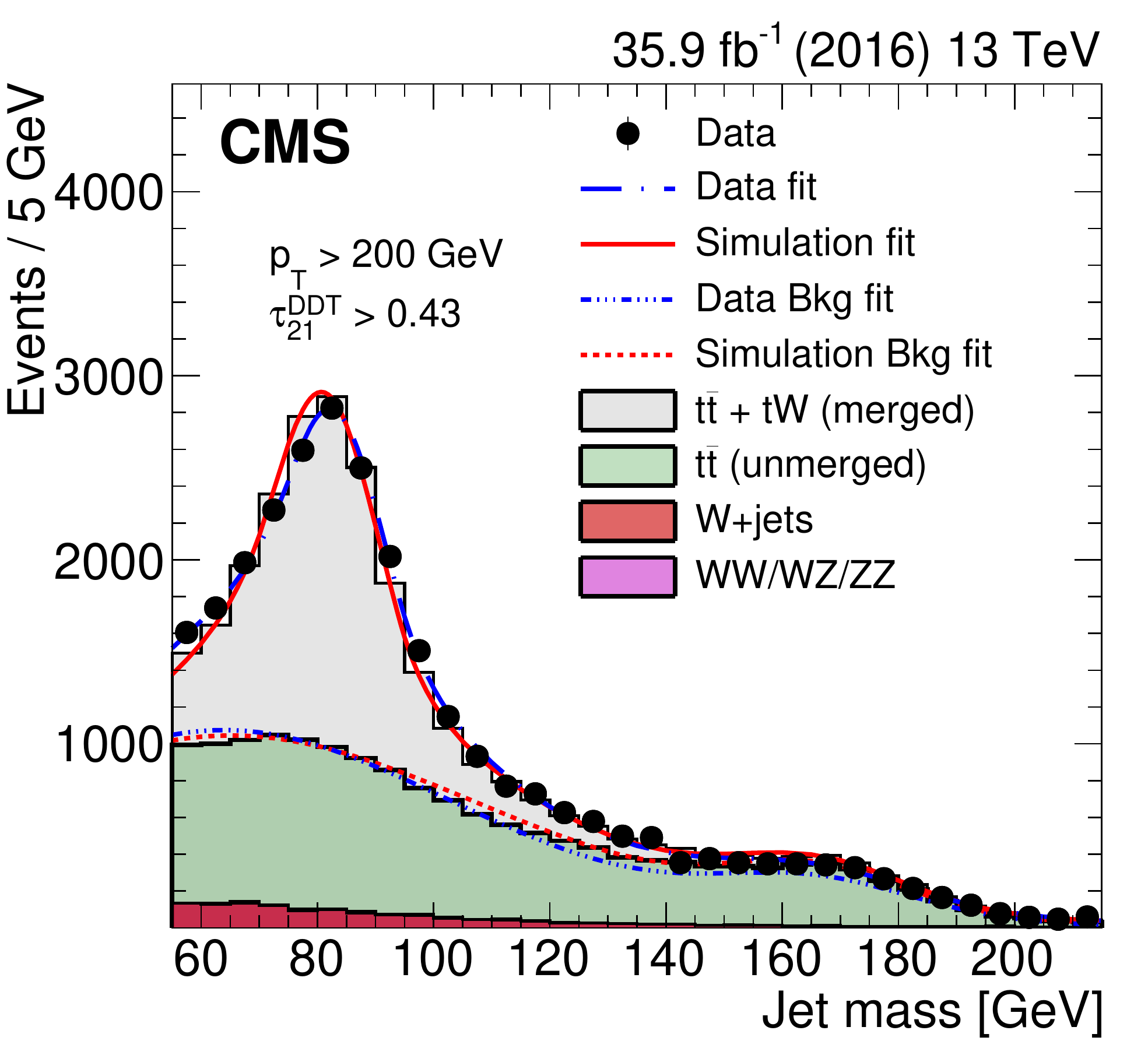}}
\vspace*{8pt}
\caption{Extraction of V-tagging scalefactors are done using a simultaneous fit in the \nsubj pass (left) and fail (right) region. Taken from Ref.~\protect\refcite{b2g18002}.\protect\label{fig:SFddt}}
\end{figure}

Correcting for mismodeling in simulation that affect the V-tagging efficiency is extremely important to not over- or under-estimate the signal yield. Therefore, dedicated corrections for efficiency, jet mass scale and jet mass resolution are derived in a dedicated semi-leptonic \ttbar-enriched region, where a large fraction of merged W boson jets are present. This is done by requiring events with one high-\PT lepton, missing energy, at least one b-tagged AK R=0.4 jet and one high-energy AK8 jet. The AK8 jet is selected as the W-jet candidate and is used to calibrate the jet tagging efficiency. Figure~\ref{fig:sltt} shows the event objects, where the semi-leptonically decaying top quark is used as a 'tag' and the hadronically decaying top as a 'probe'.
The V-tagging efficiency is measured from the PUPPI softdrop jet mass spectrum as follows. First, the mass spectrum is split into two regions: jets passing the \nsubj selection ({\it pass}) and jets failing the \nsubj selection ({\it fail}). Then, probability density functions (pdfs) which describe the distribution of fully merged W boson jets and non-W boson jets, both in the pass and in the fail region, are defined. The pdfs describing real W jets and non-W jets are added with a fraction which is left floating: The fit decides what the fraction of real W to non-W jets is in the pass and in the fail region. A simultaneous fit of pass and fail regions is then performed (using the two composite pdfs), where the fraction of real W jets in both pass and fail regions is constrained such that, if the signal efficiency in the pass region is $\epsilon_S$, the signal efficiency in the fail region is ($1-\epsilon_S$). This is done by letting the normalization of the pdf describing real W jets in the pass category be defined as the total real W boson yield in the pass and fail regions combined, multiplied by some fraction, $\epsilon_S$. The normalization of the pdf describing real W boson jets in the fail category is then the total real W boson yield in the pass and fail regions combined, multiplied by ($1-\epsilon_S$). This simultaneous fit is done to both data and simulation. The jet tagging efficiency $\epsilon_S$ is then defined from the integral under the Gaussian components of the fit, the jet mass scale from the Gaussian mean and the jet mass resolution from the Gaussian width. The ratio between the extracted values of these three parameters in data and in simulation are taken as the efficiency, jet mass scale and jet mass resolution correction factors. An example of such a simultaneous fit is shown in Figure~\ref{fig:SFddt}.
The solid blue line is the total fit to data and the solid red line is the corresponding fit to MC, for the \nsubj pass category (left), and the \nsubj fail category (right). The estimated background component is shown as dotted lines. In the fit of the fail category, a bump at around 170 GeV is observed. This is due to fully merged top jets, where the b-quark and the quarks from the W jet decay are merged into a single jet. This happens at increasing rates as a function of \PT and therefore limits the possibility of extracting V-tagging scalefactors at very high transverse momentum. Any potential differences in extracted scalefactors due to simulation of the \ttbar topology, is accounted for as a systematic uncertainty by comparing the central value of the scale factor obtained using other MC generators, like \PYTHIA{8}+\POWHEG(NLO) and \PYTHIA{8}+\MADGRAPH(LO).

\subsection{Inclusion of a search for excited quarks}
To extend the reach of the analysis, a search for excited quarks~\cite{HARRIS_2011} ${\rm q^*}$ decaying to qV was included by removing the V-tagging requirement on one of the two jets. This analysis was the very first search for excited quarks in the qV channel using 13 TeV data, and is complementary to generic dijet searches which look for ${\rm q^* \rightarrow q \bar{q}}$. No significant deviations from the  Standard Model prediction were observed in either of the searches, leading to the exclusion of non-SM resonances decaying to VV (where V=W/Z) or qV up to very high masses, $\sim3$ TeV for VV signals and $\sim5$ TeV for qV signals. Further details can be found in Ref.\protect\refcite{Sirunyan:2017acf}.

\section{A novel framework for multi-dimensional searches}
\label{sec:3}
\begin{figure}[ht!]
\centerline{\includegraphics[width=0.99\textwidth]{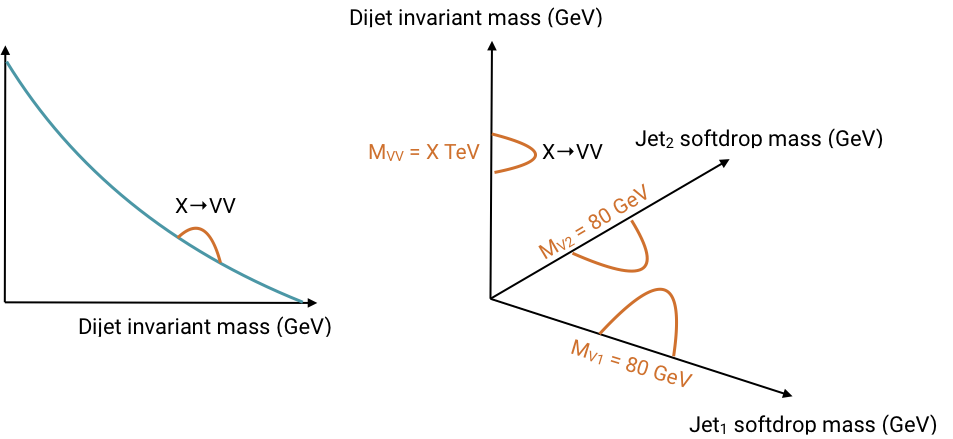}}
\vspace*{8pt}
\caption{With the new multi-dimensional fit method, rather then extracting the signal using a one-dimensional fit of$m_{\rm{jj}}$, it is extracted from the $m_{\rm{jet1}}$-$m_{\rm{jet2}}$-$m_{\rm{jj}}$ hyperplane where it peaks in all three dimensions.\protect\label{fig:3D}}
\end{figure}

No excesses had been observed in analyses of the 2015 and 2016 data sets described above. For future diboson searches, it was therefore of interest to look beyond standard signal hypotheses by establishing a method to efficiently probe a large range of different signal hypotheses. There could still be signal present in CMS data, but it might look slightly different than previously assumed. For instance, new signals could exist where the observed jet mass is slightly different than that of a W/Z boson mass. Further, these could be 4-pronged objects rather than 2-prong, which would cause the excess to vary in size depending on the analysis-specific, vector-boson tagger in use. 
In order to efficiently search for a broader range of signals, a generic framework that would allow searching for peaks anywhere in the jet mass and dijet invariant mass spectrum was developed in Ref.~\protect\refcite{b2g18002}.
Rather than selecting jets with a jet mass compatible with the W/Z boson mass and searching for resonances peaking in the dijet invariant mass, a search would be performed for resonances peaking anywhere in the hyperplane formed by the mass of each jet and their dijet invariant mass, scanning the full mass spectrum in a single analysis, as illustrated in Fig.~\ref{fig:3D}. 
This new method was developed in the context of the diboson all-hadronic search, which would allow for a straight forward comparison of the obtained results with the previous search presented in Section~\ref{sec:2}, before including searches for resonances decaying to VV (V=W/Z), VH (H=Higgs), HH, VX, VH, XX, or XY, where X and Y are new hypothetical bosons, in the same analysis. There are also expected sensitivity gains when using a multidimensional fit. One is due to a jet-mass selection no longer being needed, as the full jet mass line-shape is fitted to extract the signal. This effectively increases the signal statistics since a large fraction of the W/Z signal falls outside the mass window (20\%). Fitting the groomed-jet mass and resonance mass together also allows for the addition of nuisance parameters that simultaneously affect both, in order to fully account for the correlation between the variables. Background fluctuations are also less likely, as the signal is required to peak in three dimensions.  Finally, the background modeling would start from simulation rather than from a dijet fit to data. This allows the background shape to assume non-smooth distributions, and would allow the search to probe lower dijet invariant masses. Replacing the parametric fit by a simulation-based model also reduces the fit sensitivity to background fluctuations in the extreme tails of the dijet invariant mass spectrum. This new technique of background modeling is complex, and requires Gaussian kernels with a mean and width obtained through forward-folding rather than single points in order to model the QCD multijet distribution. The 2016 data set was re-analyzed in order to directly compare the sensitivity between the previous method described in Section~\ref{sec:2} and the multi-dimensional technique summarized here, and was also applied when performing the first analysis of the data collected in 2017.

\subsection{Decorrelating V-tagging from transverse momentum and mass}
\begin{figure}[ht]
\centerline{
\includegraphics[width=0.49\textwidth]{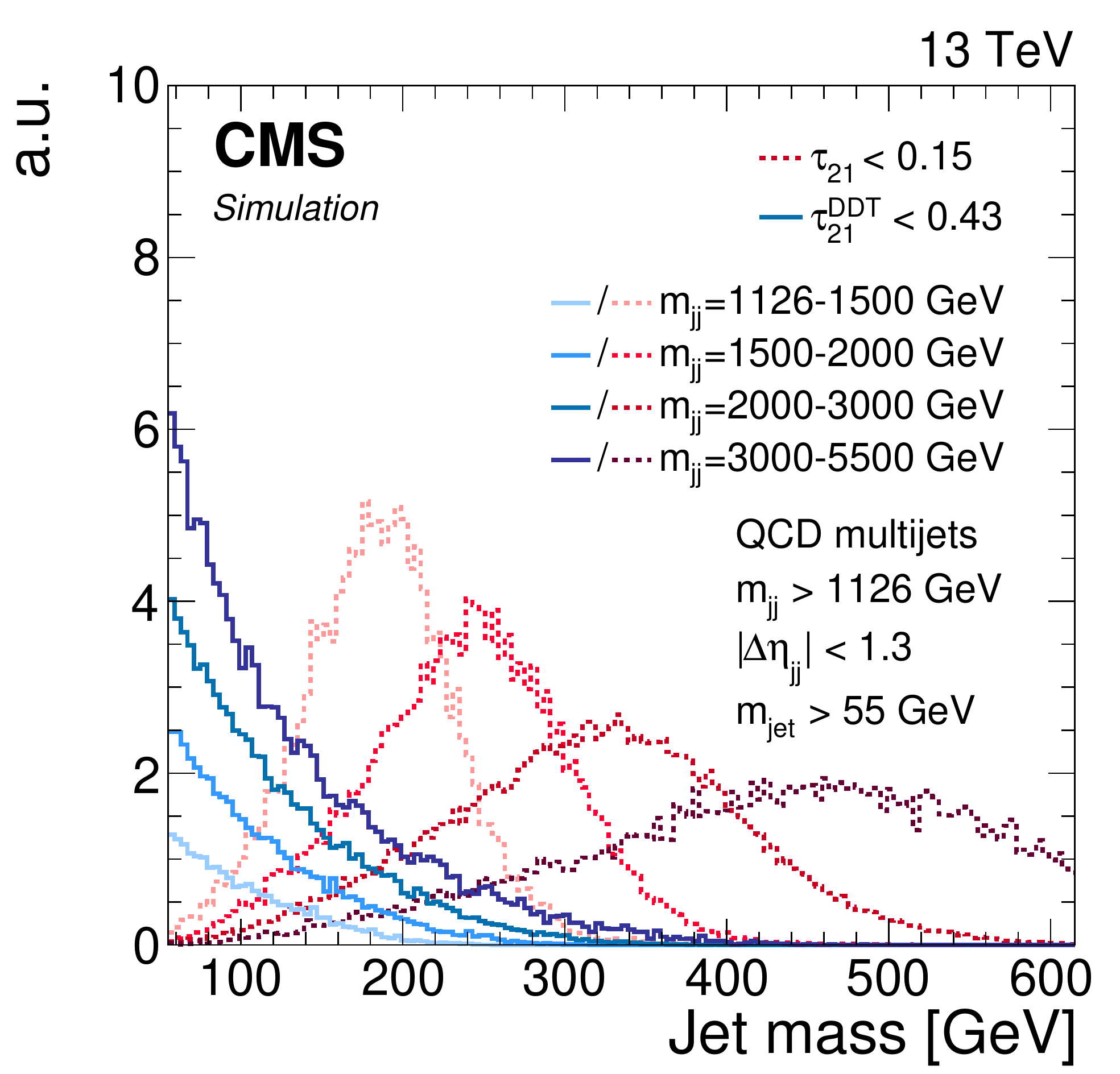}}
\vspace*{8pt}
\caption{The jet mass distribution observed in QCD simulation after applying a selection on \nsubj (dotted lines) or \ddt (solid lines) in bins of dijet invariant mass. Taken from Ref~\protect\refcite{b2g18002}.\protect\label{DDT}}
\end{figure}

After applying the same pre-selections as discussed in Section~\ref{sec:1}, the jets are randomly sorted (as to which jet mass value is assigned to the y-axis and which jet mass value is assigned to the x-axis) such that the two jet mass distributions have the same mean and spread. No mass selection is applied as the jet masses now are variables of interest. The vector-boson tagging, in this case, therefore only includes a selection on the shape variable N-subjettiness. A mass- and \PT-decorrelated version of \nsubj intended to remove mass sculpting was developed in the context of this analysis since modeling of the background in a very wide range of mass and transverse momenta is required. It is desirable that the QCD background spectrum is minimally sculpted as a function of \PT and mass, such that the background shapes in all regions are similar to one another and remains smoothly falling in all three analysis dimensions. In order to ensure minimal sculpting, the \nsubj variable is decorrelated from the jet softdrop mass and the jet \PT using the procedure following the {\it Designing Decorrelated Taggers}~\cite{ddt} methodology. This decorrelation is performed by flattening the \nsubj profile dependence on $\rho'= log(m^2/p_T/\mu)$, where $\mu$ = 1 GeV.
A linear transformation is defined as
\begin{equation}
\label{eq:ddt}
\tau_{21}^{DDT}=\tau_{21}-M\times\rho',
\end{equation}
where the slope M is fitted from the linear part of the \nsubj profile versus $\rho'$. Figure~\ref{DDT} shows the jet mass distribution for the QCD background in bins of dijet invariant mass after a cut is applied on \nsubj (dotted lines) or \ddt (solid lines). 
The background remains smoothly falling after a selection on \ddt, whereas a selection on \nsubj significantly sculpts the background.
In addition to yielding better decorrelation, the \ddt tagger also has a significantly higher signal efficiency for a given mistag rate for this search as it is a function of both \nsubj and the ratio of jet mass and \PT.

\subsection{Modeling through forward folding}
\label{sec:4}
In order to better constrain the background at high dijet invariant masses, the background is modeled starting from simulation rather than using the parametric fit described in Section~\ref{sec:1}. The QCD simulation is limited by statistics, especially in the most extreme mass bins, and a forward folding approach is therefore used in order to ensure a smooth background template across all bins. The jet mass is correlated with the dijet mass and is modeled conditionally (in bins of dijet invariant mass) as two-dimensional templates, and the dijet invariant mass is modeled as a one-dimensional template. The forward-folding approach consists of starting from particle-level information, like jet mass and/or dijet invariant mass, and then smearing and scaling each event based on the jet mass resolution and scale using a Gaussian kernel (2D Gaussian in the case of a two-dimensional template, and 1D in the case of a one-dimensional template). Therefore, rather than filling a 1D/2D histogram with one event, the histograms are filled with smooth Gaussian kernels ensuring a valid probability density function in all mass bins, following what was proposed in Ref.~\protect\refcite{Cranmer:2000du}. The three templates are finally multiplied together forming one joint pdf to model the three-dimensional hyperplane spanned by the jet masses and dijet invariant mass. Nuisance parameters with large uncertainties are added to allow the background template to change in order to best fit the observed data. These include adding alternative templates derived using different MC generators, like \HERWIG and \MADGRAPH, effectively incorporating all known background variations into the final fit.
\vskip 1.em
\noindent The signal is modeled separately for all three analysis dimensions and no correlation between jet mass and dijet invariant mass is assumed (the signal is resonant both in jet mass and dijet invariant mass and therefore localized).
\begin{figure}[ht]
\centerline{
\includegraphics[width=0.49\textwidth]{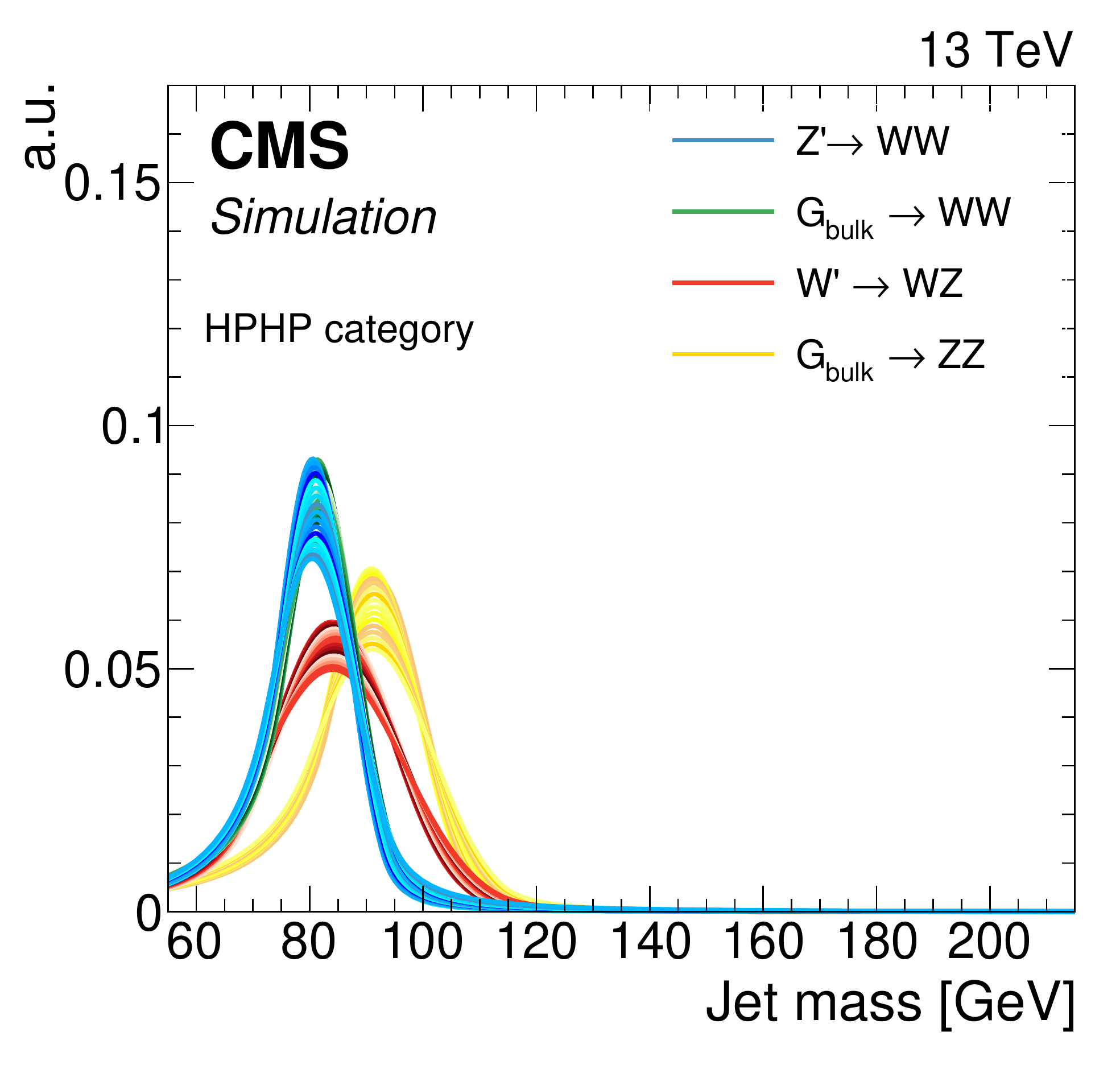}}
\vspace*{8pt}
\caption{Jet mass shapes for different signal hypotheses and a range of different resonance masses between 1.2 and 5.2 TeV. Taken from Ref.~\protect\refcite{b2g18002}.\protect\label{mjetsignal}}
\end{figure}
 The parametrization is done by fitting the simulated jet mass and dijet invariant mass for each signal resonance mass point, and then parametrizing these as a function of resonance mass. The resonance mass shapes are similar to those in Figure~\ref{fig4}, and the jet mass shape for different signal hypotheses is shown in Figure~\ref{mjetsignal}. The jet mass is fitted with a double-sided Crystal Ball function, and the jet mass scale and resolution are extracted from the double Crystal Ball mean and width. 

\subsection{First observation of SM V(${\it q\bar{q}}$) in diboson analyses}
\begin{figure}[ht!]
\centerline{
\includegraphics[width=0.49\textwidth]{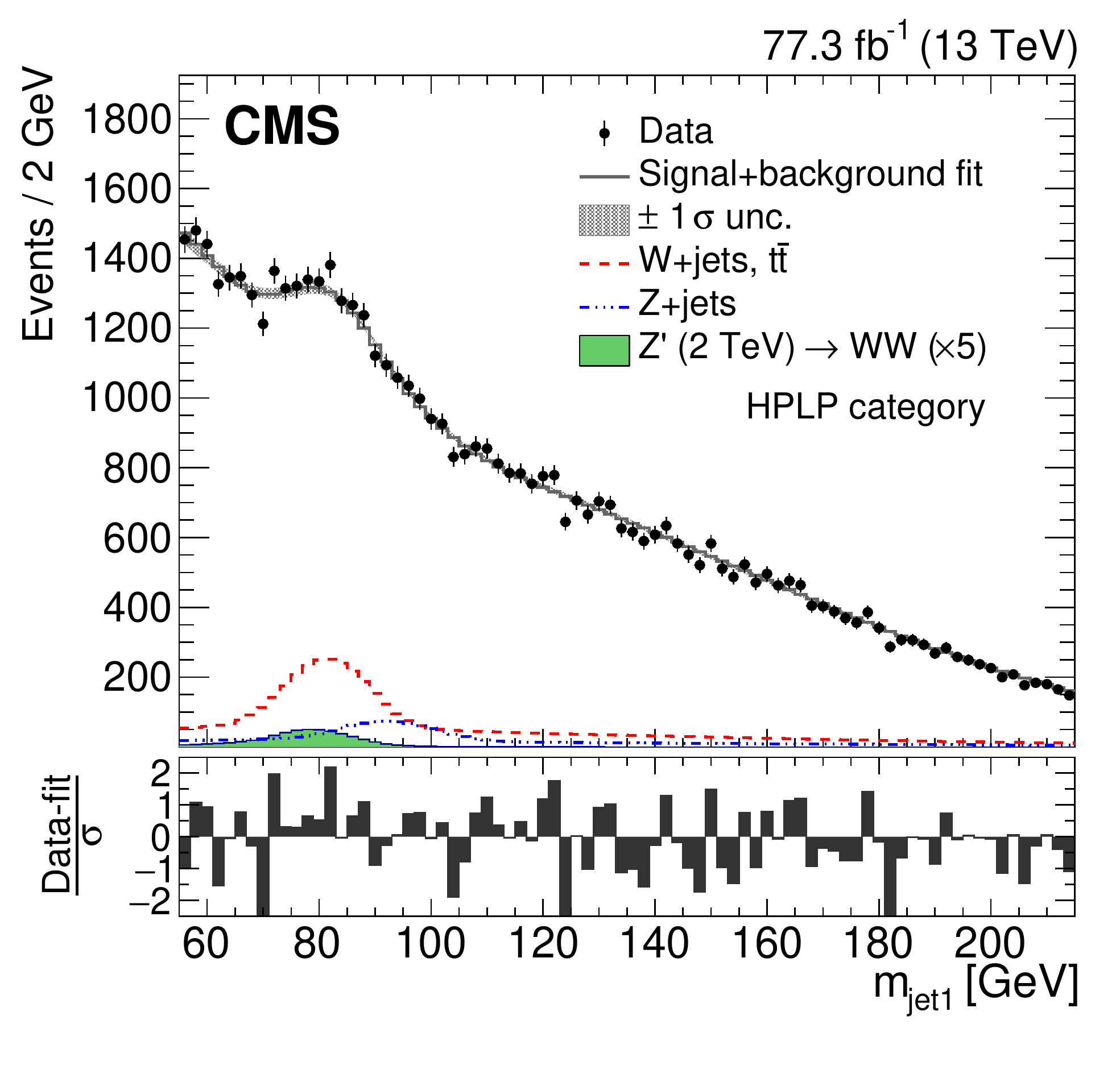}
\includegraphics[width=0.49\textwidth]{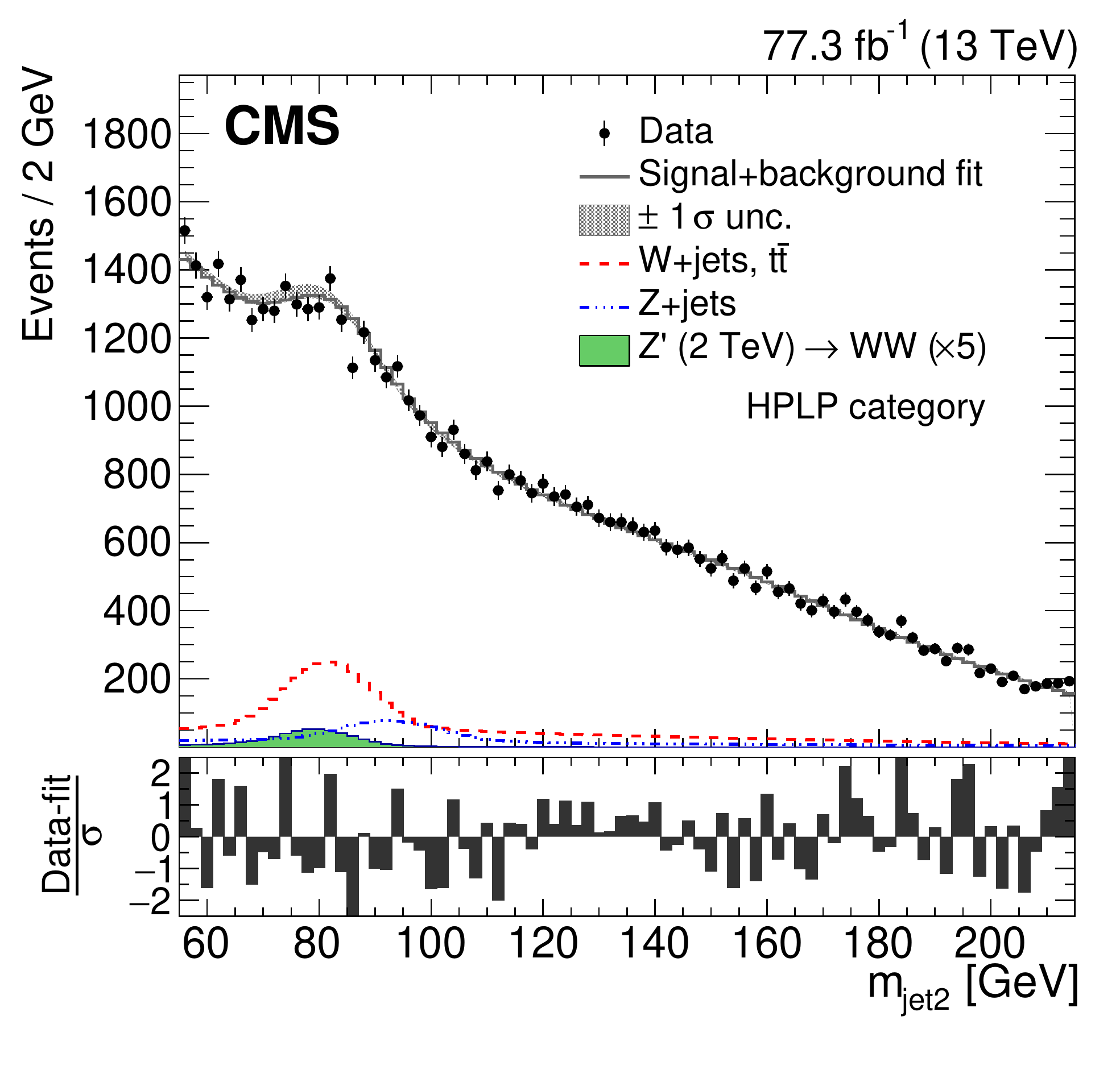}}
\centerline{
\includegraphics[width=0.49\textwidth]{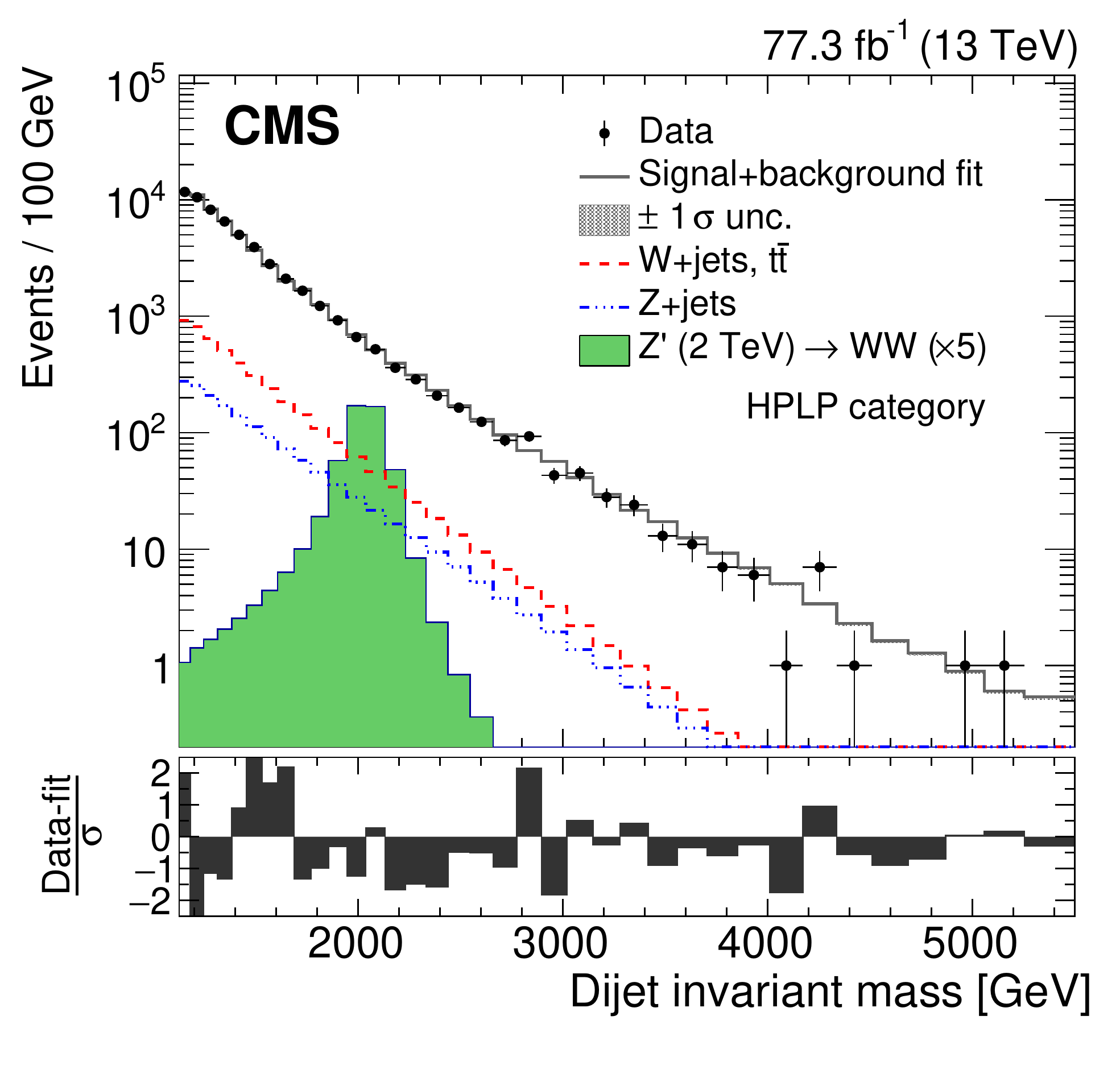}}
\vspace*{8pt}
\caption{Comparison between the fitted result and data distributions for the jet mass of the two jets (top left and right) and the dijet invariant mass (bottom). An example of a signal distribution is overlaid, using an arbitrary normalization. Two beautiful peaks from Z(qq)+jets and W(qq)+jets are measured for the first time in diboson analyses. Taken from Ref.~\protect\refcite{b2g18002}.\protect\label{fig:PFit}}
\end{figure}

The final fitted result and the data distributions projected onto the three dimensions of interest are shown in Figure~\ref{fig:PFit}. Two beautiful peaks from the Standard Model Z(qq)+jets and W(qq)+jets background are observed. This is the first time these SM backgrounds have ever been measured in a diboson analysis. Their extracted cross sections are found to be compatible with the SM expectation~\cite{b2g18002}, an observation made possible due to the nature of the fit and the optimized vector-boson tagging algorithm. The two peaks also allow uncertainties affecting the signal yield to be constrained, leading to a better analysis sensitivity.

\subsection{Sensitivity increase using multidimensional techniques}

\begin{figure}[ht]
\centerline{\includegraphics[width=0.49\textwidth]{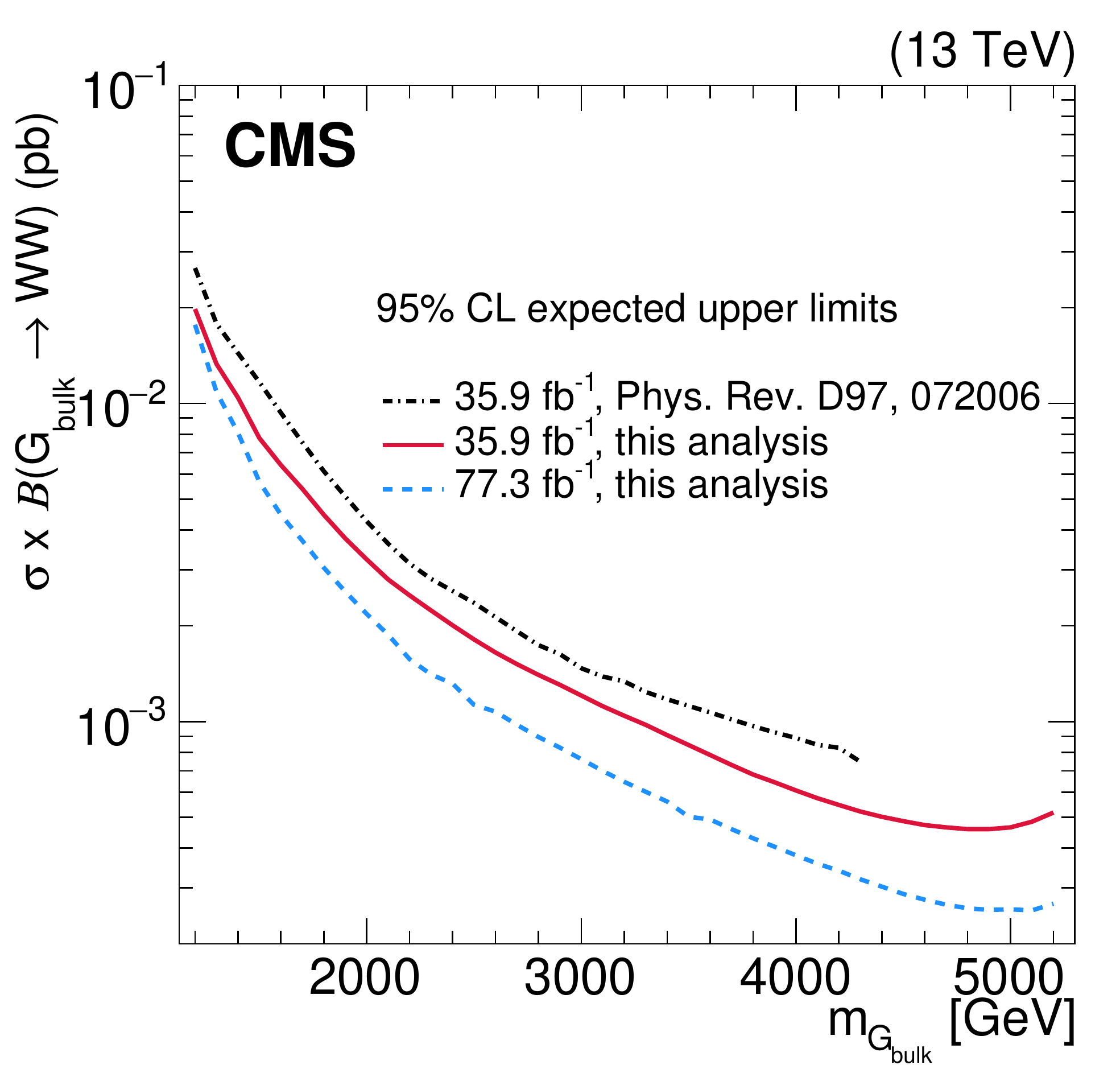}}
\vspace*{8pt}
\caption{Expected limits for a Bulk $G\rightarrow WW$ signal obtained using the multi-dimensional fit method presented here (red line), compared to the result obtained using previous methods (black dotted line)~\cite{Sirunyan:2017acf}. The final limit obtained when combining data collected in 2016 and 2017 is also shown (blue dotted line). Taken from Ref.~\protect\refcite{b2g18002}.\protect\label{fig:limits}}
\end{figure}

The obtained expected upper limits using the multi-dimensional fit method introduced in Ref.~\refcite{b2g18002}, can be compared to those obtained in the previous search described in Section~\ref{sec:2} using the same data set, Ref.~\cite{Sirunyan:2017acf}, in order to estimate whether there is a sensitivity gain in using the new method. Figure~\ref{fig:limits} shows the expected limits based on analyses of the data collected in 2016, either using the fit method presented here, or using the previous one-dimensional method. A 20-30\% improvement in sensitivity is obtained using the multidimensional fit method, and about a 35-40\% improvement is obtained when combining the two data sets with respect to the individual results.
The analysis of the 2016 and 2017 data sets using the multidimensional fit are the most sensitive to date in the diboson channel and are as sensitive as the full 2016 combination of diboson and leptonic analysis channels in Refs.\protect\refcite{Sirunyan_2019,Aaboud_2018}, by the CMS and the ATLAS collaboration, respectively. Analysis of the full dataset collected by the CMS experiment at $\sqrts=13$~TeV is ongoing, including searches for VV, VH, and HH using improved taggers, and combined in one common analysis, and is expected to improve the sensitivity of the searches even further.

\section{Summary}
In this review article, the CMS search program for heavy resonances decaying to WW, ZZ, or WZ boson pairs in final states with boosted jets using data collected at $\sqrts=13$~TeV has been summarized. Improvements of vector-boson tagging methods and search techniques benefiting the analyses have been emphasized. Collectively, the improvements summarized above have resulted in the most sensitive limits to date in all-hadronic diboson resonance searches.

\section*{Acknowledgement}
I thank all my colleagues in the CMS experiment for the excellent work in making these results possible, and gratefully acknowledge enlightening discussions with those who conducted diboson resonance searches during 13 TeV data taking. This especially includes Jennifer Ngadiuba, Daniela Sch{\"a}fer, Andreas Hinzmann, Ben Kilminster, Clemens Lange, and Maurizio Pierini. 
\bibliographystyle{ws-mpla}
\bibliography{bibliography}

\end{document}